\documentclass[conf]{new-aiaa}
\usepackage[utf8]{inputenc}

\usepackage{graphicx}
\usepackage{amsmath}
\usepackage[version=4]{mhchem}
\usepackage{siunitx}
\usepackage{longtable,tabularx}
\usepackage{comment}
\usepackage{subcaption}

\usepackage{amsthm}

\newtheorem{lemma}{Lemma}
\newtheorem{assumption}{Assumption}

\title{Contraction-based Neural Control for Cooperative Aerial Payload Transportation with Variable-length Cables\footnote{
This manuscript is a preprint submitted to the American Institute of Aeronautics and Astronautics (AIAA) for publication. Copyright may be transferred to AIAA upon publication. This version has not undergone AIAA copyediting, typesetting, or formatting.
}}

\author{Yi Lok Lo\footnote{Graduate Student, enoch.lo@mail.utoronto.ca}, Longhao Qian\footnote{Post-doctoral Fellow, longhao.qian@mail.utoronto.ca} and Hugh H.T. Liu\footnote{Professor, hugh.liu@utoronto.ca, Associate Fellow of AIAA}}
\affil{University of Toronto Institute For Aerospace Studies (UTIAS), Toronto, ON, Canada M3H 5T6}

\begin{document}

\maketitle

\begin{abstract}
This paper presents a novel neural nonlinear control framework for a multi-drone slung payload system with variable-length cables and a rigid-body payload. The equations of motion are formulated into a decoupled structure, where the payload and cable length dynamics are governed by independent control channels, facilitating modularized controller design on reduced-order subsystems. A neural control contraction metric (CCM) controller and a neural feedback controller are jointly trained to enforce contraction conditions for the payload subsystem. Separately, a cable length control law is derived that exploits the variable-length degree of freedom for obstacle avoidance. Numerical simulations demonstrate trajectory tracking of a rigid-body payload and gate traversal capabilities of the overall system under the proposed control framework.
\end{abstract}
\vspace{-0.3cm}
\section*{Nomenclature}

{\renewcommand\arraystretch{1.0}
\noindent\begin{longtable*}{@{}l @{\quad=\quad} l@{}}

$c_j$                   & quadrotor-to-payload mass ratio, 
                          $m_j / m_p$ \\
$\boldsymbol{e}_1,\boldsymbol{e}_2,\boldsymbol{e}_3$  & Euclidean basis vectors $[1,0,0]^\top$, $[0,1,0]^\top$, $[0,0,1]^\top$   \\                     
$f_{||,j}$              & magnitude of lift force component 
                          parallel to $j^{th}$ cable, N \\
${\boldsymbol{f}}_{\bot,j}$ 
                        & lift force component 
                          perpendicular to $j^{th}$ cable, 
                          N \\
$\boldsymbol{f}_{L,j}$  & lift force of $j^{th}$ quadrotor 
                          expressed in the inertial frame, N \\
$f_{T,j}$               & tension in $j^{th}$ cable, N \\
$\boldsymbol{g}_I$      & gravitational acceleration vector 
                          in the inertial frame, $\begin{bmatrix}
                              0,0,-9.81
                          \end{bmatrix}^\top$, m/s$^2$ \\
$\boldsymbol{I}_n$, $\boldsymbol{0}_{n\times m}$      & $n \times n$ identity matrix and $n \times m$ zero matrix, respectively \\
$j$                     & quadrotor and cable index, $j = 1,2, \dots, N$ \\
$\boldsymbol{J}_j$      & moment of inertia matrix of $j^{th}$ 
                          quadrotor, kg$\cdot$m$^2$ \\
$\boldsymbol{J}$      & moment of inertia matrix of rigid-body 
                          payload, kg$\cdot$m$^2$ \\
$l_j$                   & length of $j^{th}$ cable, m \\
$m_j$                   & mass of $j^{th}$ quadrotor, kg \\
$m_p$                   & mass of rigid-body payload, kg \\
$N$                     & total number of quadrotors \\
$\boldsymbol{n}_j$      & unit vector representing $j^{th}$ cable direction\\
$\boldsymbol{R}_{Ij}$   & rotation matrix from $j^{th}$ quadrotor 
                          body-fixed frame to the inertial frame \\
$\boldsymbol{R}_{IP}$   & rotation matrix from the payload body-fixed frame to the inertial frame \\
$\boldsymbol{r}_j$      & 2D horizontal projection of $j^{th}$ cable directional vector $\boldsymbol{n}_j$ \\
$\boldsymbol{t}_j$      & position of $j^{th}$ cable attachment 
                          point in payload body frame, m \\
$\boldsymbol{v}_p$      & inertial velocity of payload, m/s \\
$\boldsymbol{v}_{j,I}$  & inertial velocity of $j^{th}$ 
                          quadrotor, m/s \\
$\boldsymbol{v}_j$      & 2D horizonal projection of normalized cable swing velocity of $j^{th}$ cable \\
$\boldsymbol{x}_p$      & inertial position of payload, m \\
$\boldsymbol{x}_j$      & inertial position of $j^{th}$ 
                          quadrotor, m \\
$\boldsymbol{Y}_j$      & inertia-scaled mass matrix, $m_j \boldsymbol{J}^{-1}$ \\
$\boldsymbol{z}_j$      & 2D independent cable swing control 
                          input \\
$\boldsymbol{\tau}_j$   & control torque applied to the $j^{th}$ quadrotor, N$\cdot$m \\
$\boldsymbol{\omega}_p$ & angular velocity of payload, 
                          rad/s \\
$\boldsymbol{\omega}_j$ & angular velocity of $j^{th}$ 
                          quadrotor, rad/s \\
$(\cdot)^\times$            & skew-symmetric matrix of a 
                              vector \\
$(\cdot)^\vee$              & vee operator, inverse of 
                              $(\cdot)^\times$ \\
$||\cdot||$                 & Euclidean norm \\
$\langle\boldsymbol{A}\rangle$ 
                            & symmetric part of matrix $\boldsymbol{A}$, 
                              $(\boldsymbol{A} + 
                              \boldsymbol{A}^T)/2$ \\
$\mathrm{diag}(\cdot)$      & block diagonal matrix 
                              operator \\
$\mathrm{vec}(\cdot)$       & row-major vectorization of a matrix \\
$\prec,\succ$               & matrix comparisons, $\boldsymbol{A}\succ\boldsymbol{B}$ means $\boldsymbol{A}-\boldsymbol{B}$ is positive definite for symmetrical matrices $\boldsymbol{A}$,$\boldsymbol{B}$ \\
$\partial_{\boldsymbol{f}}\boldsymbol{A}(\boldsymbol{x})$  & Lie-derivative of $\boldsymbol{A}(\boldsymbol{x})$ along vector field $\boldsymbol{f}$, $\sum_i\frac{\partial\boldsymbol{A}}{\partial x_i}f_i$, where $x_i$,$f_i$ denotes $i^{th}$ element of $\boldsymbol{x}$,$\boldsymbol{f}$

\end{longtable*}}

\section{Introduction}

\lettrine{M}{ulti-drone} cooperative transportation systems have attracted significant research interest due to their ability to carry payloads beyond the capacity of a single drone, while offering improved fault tolerance and enhanced maneuverability~\cite{Qian2019,AlLawati2025}. Compared to single-drone configurations, multi-drone systems provide richer actuation authority over the payload pose, enabling simultaneous control of both translational and rotational motion of rigid-body payloads suspended by cables~\cite{Wahba2024}. The dynamics of multi-drone slung payload systems are highly nonlinear and underactuated, posing fundamental challenges for controller design and stability analysis. The work in \cite{Wu2014} pioneered geometric control for multiple quadrotors transporting a rigid-body load, proving almost global stability under simplified assumptions. A robust control architecture is developed in \cite{Qian2022} for a similar system, proving Lyapunov stability of the cascaded closed-loop system and validating the approach experimentally. 

The introduction of variable-length cables provides an additional degree of freedom that can be exploited. The geometric control and differential flatness properties are derived in \cite{Zeng2019}, which underpins trajectory generation for single-drone variable-length cable systems. Anti-swing control strategies are proposed that effectively suppresses swing while maintaining trajectory tracking performance \cite{Yang2022,Huang2023,Yu2023}. Experimental validations are also performed in \cite{Li2023,Prajapati2022,Yu2025a}, showing improvements for outdoor aerial transportation using active length adjustment. An online cable length optimization framework is also proposed in \cite{Yu2026} that utilizes the extra degree of freedom to facilitate smooth navigation through narrow spaces for single-drone systems. 

Incorporating variable cable lengths into a multi-drone payload transportation system is therefore proposed in this work, as the actuated cable lengths can reshape the size of the overall system, allowing aerial transportation in a cluttered environment. The work in \cite{Li2020} proposed a feedback linearization-based control framework for variable aerial cable towed systems, but it requires the quadrotors to maintain specific configurations to ensure the whole system is over-actuated, limiting flexibility under general formation changes. A distributed model predictive control (MPC) strategy is also proposed in \cite{Tartaglione2017}, taking into account the operational constraints such as collision avoidance and control saturation. 

Despite this growing body of literature, a controller with formal stability guarantees for the variable-length multi-drone rigid-body payload system has yet to be established. The high dimensionality of the state space, combined with the complex nonlinear coupling introduced by the rigid-body payload attitude dynamics and the variable-length cable kinematics, makes analytical controller design particularly challenging. MPC-based approaches could in principle handle the system constraints, but the associated online optimization is computationally prohibitive at the state space dimension required for this system. It therefore remains an open problem to find a controller that simultaneously provides formal stability guarantees, handles the high-dimensional nonlinear dynamics, and admits a lightweight real-time implementation without online optimization.

In recent years, learning-based controller design has gained significant traction as a means of handling complex dynamics. In particular, control contraction metrics (CCMs) are widely recognized as a constructive nonlinear controller design framework since the seminal works in \cite{Manchester2017a,Lohmiller1998}. Since then, several learning-based CCM frameworks have been proposed, providing interpretable and certifiable stability guarantees \cite{Tsukamoto2020,Sun2021}. Importantly, the contraction conditions can be enforced during offline training, shifting the computational burden away from online implementation and enabling real-time deployment without runtime optimization \cite{Tsukamoto2021}. 

The primary contributions of this work are as follows:
\begin{enumerate}
    \item A decoupled formulation of the multi-drone variable-length rigid-body slung payload system is derived using the Newton-Euler method, where the payload movement and cable length dynamics are governed by independent control channels, enabling modular controller design on reduced-order subsystems.
    \item A neural CCM controller is trained on the payload subsystem, utilizing deep learning to design controllers for high-dimensional nonlinear systems based on contraction theory. 
    \item The variable-length degree of freedom is exploited for obstacle avoidance through height-constrained environments via cable length profiling. Numerical simulations of the complete system are performed, demonstrating successful trajectory tracking and obstacle avoidance capabilities using the proposed controller framework.
\end{enumerate}

\section{System Modelling}

\begin{figure}[htbp]
    \centering
    \includegraphics[width=0.62\textwidth]{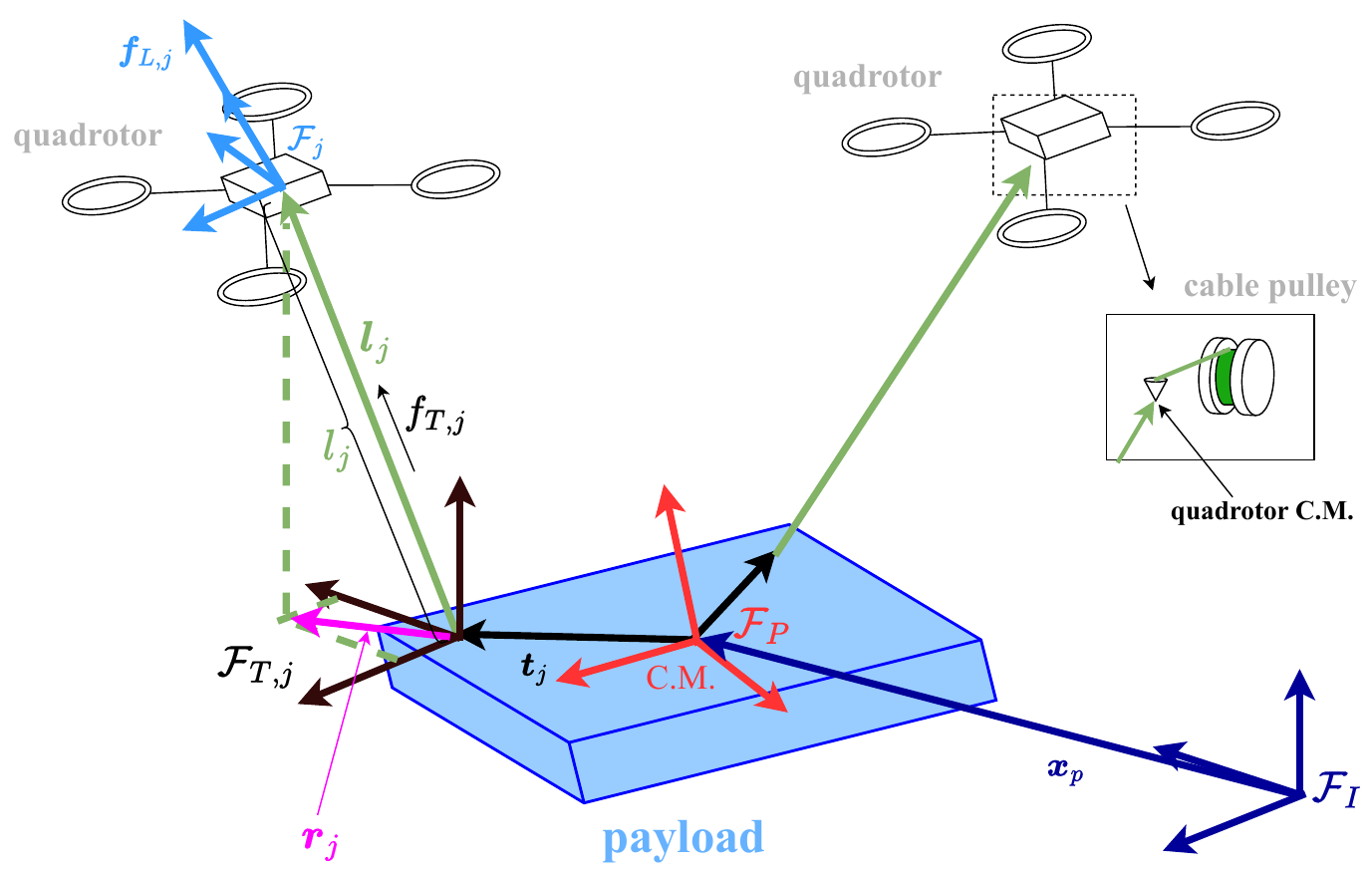}
    \caption{Geometry of the system}
    \label{fig: system_geometry}
\end{figure}
\vspace{-0.1cm}

\subsection{Reference Frames and Kinematics}\label{sec: kinematics}

The geometry of the system is shown in Fig. \ref{fig: system_geometry}. $\mathcal{F}_I$ is the inertial frame. $\mathcal{F}_P$ is the body-fixed frame on the center of mass (C.M.) of the payload. Also, $\mathcal{F}_{T,j}$ is defined as the tether frame of the $j^{th}$ cable, with its origin fixed at the cable attachment point on the payload. This frame only translate with the payload, and the rotation matrix between $\mathcal{F}_{T,j}$ and $\mathcal{F}_I$ is always identity. Lastly, $\mathcal{F}_{j}$ is defined as the body-fixed frame on the $j^{th}$ quadrotor. 

\begin{assumption}\label{assum: system}
    The following assumptions on the system are made. First, the cables are assumed to be taut, such that the tension in the cables $f_{T,j}$ are positive at all times. Moreover, the $j^{th}$ cable is assumed to be attached at the C.M. of the $j^{th}$ quadrotor, such that the attitude dynamics of the drone are decoupled from the payload system. Lastly, the cable vector from $\mathcal{F}_{T,j}$ to $\mathcal{F}_{j}$ expressed in $\mathcal{F}_{T,j}$ has positive $z$-component, meaning it always points above the $xy$-plane of $\mathcal{F}_{T,j}$. 
\end{assumption}

The kinematics of the system is modelled under Assumption~\ref{assum: system}. According to Fig. \ref{fig: system_geometry}, the cable vector from the attachment point on the payload to that of the quadrotor $\boldsymbol{l}_j\in\mathbb R^{3\times 1}$ and its associated directional vector $\boldsymbol{n}_j\in\mathbb R^{3\times 1}$ are defined as
\begin{equation}\label{eq: def_cable_vector}
\boldsymbol{l}_j = l_j \boldsymbol{n}_j = \begin{bmatrix}
l_j \boldsymbol{r}_j\\
l_j \sqrt{1 - \boldsymbol{r}_j^T\boldsymbol{r}_j}
\end{bmatrix} = l_j \begin{bmatrix}
 \boldsymbol{r}_j\\
 \sqrt{1 - \boldsymbol{r}^T_j\boldsymbol{r}_j}
\end{bmatrix}, \quad \boldsymbol{n}_j =  \begin{bmatrix}
 \boldsymbol{r}_j\\
 \sqrt{1 - \boldsymbol{r}^T_j\boldsymbol{r}_j}
\end{bmatrix}.
\end{equation}
From the above equation, $l_j$ is the length of the $j^{th}$ cable, and $\boldsymbol{r}_j$ is the normalized relative position of the quadrotor, which represents the 2-D offset
of the quadrotor when the cable is 1m long. $\dot{\boldsymbol{r}}_j=\boldsymbol{v}_j$ is the normalized quadrotor relative
speed, where the relationship between $\boldsymbol{n}_j$ and $\boldsymbol{v}_j$ can be obtained by differentiating Eq. \eqref{eq: def_cable_vector} with respect to time. 
\begin{equation}\label{eq: B_property}
\dot{\boldsymbol{n}}_j =  \begin{bmatrix}
 \dot{\boldsymbol{r}}_j\\
 -\frac{\boldsymbol{r}^T_j\boldsymbol{v}_j }{\sqrt{1-\boldsymbol{r}^T_j\boldsymbol{r}_j}}
\end{bmatrix}=\boldsymbol{B}_j\boldsymbol{v}_j, \quad
\boldsymbol{B}_j = \begin{bmatrix}
\boldsymbol{I}_{2}\\
-\frac{\boldsymbol{r}^T_j}{\sqrt{1-\boldsymbol{r}^T_j\boldsymbol{r}_j}}
\end{bmatrix}, \quad \dot{\boldsymbol{B}}_j = \begin{bmatrix}
\boldsymbol{0}_{2 \times 2}\\
-\frac{(1 - \boldsymbol{r}^T_j\boldsymbol{r}_j)\boldsymbol{v}^T_j + (\boldsymbol{v}^T_j  \boldsymbol{r}_j)\boldsymbol{r}^T_j}{(1-\boldsymbol{r}^T_j\boldsymbol{r}_j)^{3/2}}
\end{bmatrix}
\end{equation}
Therefore, the cable vector $\boldsymbol{l}_j$ obeys the following kinematics:
\begin{equation}\label{eq: cable_kinematics}
\dot{\boldsymbol{l}}_j = \dot{l}_j\boldsymbol{n}_j + l_j \boldsymbol{B}_j\boldsymbol{v}_j, \quad \ddot{\boldsymbol{l}}_j = \ddot{l}_j\boldsymbol{n}_j + l_j \boldsymbol{B}_j\dot{\boldsymbol{v}}_j + 2 \dot{l}_j\boldsymbol{B}_j\boldsymbol{v}_j + l_j\dot{\boldsymbol{B}}_j\boldsymbol{v}_j.   
\end{equation}

\begin{lemma}\label{lem: B_property}
The following properties of the $\boldsymbol{B}_j$ matrix are true:
\begin{enumerate}
    \item $\boldsymbol{B}^T_j\boldsymbol{n}_j = \boldsymbol{0}_{2 \times 1}$ and  $ \boldsymbol{B}^T_j$ has full row rank;
    \item $\boldsymbol{B}_j( \boldsymbol{B}^T_j  \boldsymbol{B}_j)^{-1}  \boldsymbol{B}^T_j = \boldsymbol{I}_{3} -\boldsymbol{n}_j\boldsymbol{n}^T_j$.
\end{enumerate}
\begin{proof}
See Appendix~\ref{app: B_property}.  
\end{proof}
\end{lemma}

\subsection{Normalized Control Forces}

The control forces of the system are the lift force of the $j^{th}$ quadrotor $\boldsymbol{f}_{L,j} \in \mathbb{R}^{3 \times 1}$ defined in $\mathcal{F}_I$, and the tension of the $j^{th}$ cable $f_{T,j} \in \mathbb{R}$. Each lift force vector $\boldsymbol{f}_{L,j}$ is decomposed into 2 components, such that $\boldsymbol{f}_{L,j}=f_{||,j}\boldsymbol{n}_j+\boldsymbol{f}_{\perp,j}$. The normalized lift forces and cable tension are then defined as follows:
\begin{equation}\label{eq: normalized_control}
\begin{aligned}
& \hat{\boldsymbol{f}}_{\bot,j} = (\boldsymbol{I}_{3} -\boldsymbol{n}_j\boldsymbol{n}^T_j)\frac{\boldsymbol{f}_{L,j}}{m_j}, \quad \hat{f}_{||,j} = \frac{\boldsymbol{n}_j^T\boldsymbol{f}_{L,j}}{m_j}, \quad \hat{f}_{T,j} = \frac{f_{T,j}}{m_j}.
\end{aligned}    
\end{equation}
According to Lemma \ref{lem: B_property}, $\boldsymbol{B}_j$ has full column rank and spans the subspace perpendicular to $\boldsymbol{n}_j$. Hence, $ \hat{\boldsymbol{f}}_{\bot,j}$ can be written as:
\begin{equation}\label{eq: normalized_control_f_bot}
\begin{aligned}
&    \hat{\boldsymbol{f}}_{\bot,j} = l_j\boldsymbol{B}_j\boldsymbol{z}_j
\end{aligned}    
\end{equation}
where $\boldsymbol{z}_j \in \mathbb{R}^{2 \times 1}$ is the 2-dimensional independent control. A projection matrix $\boldsymbol{\Xi}_j \in \mathbb{R}^{2 \times 3}$ can also be obtained 
\begin{equation}\label{eq: projection_matrix}
\boldsymbol{\Xi}_j = \frac{1}{l_j} (\boldsymbol{B}_j^T\boldsymbol{B}_j)^{-1}\boldsymbol{B}_j^T,
\end{equation}
such that 
\begin{equation}\label{eq: normalized_bot_control_property}
     \boldsymbol{\Xi}_j   \hat{\boldsymbol{f}}_{\bot,j} =   \boldsymbol{\Xi}_j l_j\boldsymbol{B}_j\boldsymbol{z}_j =\boldsymbol{z}_j. 
\end{equation}

\subsection{Equations of Motion}

Systems of equations of motion are developed in this section. First, using Newton-Euler method, the equations of motion of each quadrotor are:
\begin{equation}\label{eq: inertial_accel}
\dot{\boldsymbol{v}}_{j, I} = \frac{1}{m_j} (\boldsymbol{f}_{L, j} -\boldsymbol{n}_j f_{T,j} ) + \boldsymbol{g}_I.
\end{equation}
Then, the equation of motion of the payload can be obtained as
\begin{equation}\label{eq: payload_accel}
\begin{aligned}
 & \dot{\boldsymbol{v}}_{p} =  \frac{1}{m_p} (\sum_{i=1}^{N} \boldsymbol{n}_i f_{T,i})  + \boldsymbol{g}_I =   \sum_{i=1}^{N} c_i\boldsymbol{n}_i \hat{f}_{T,i} + \boldsymbol{g}_I, \\
 & \dot{\boldsymbol{\omega}}_p = \boldsymbol{J}^{-1}(\sum_{i=1}^{N} \boldsymbol{t}_i^{\times}\boldsymbol{R}_{PI}\boldsymbol{n}_i f_{T,i} - \boldsymbol{\omega}_p^{\times}\boldsymbol{J}\boldsymbol{\omega}_p) = \sum_{i=1}^{N} \boldsymbol{Y}_i\boldsymbol{t}_i^{\times}\boldsymbol{R}_{PI}\boldsymbol{n}_i \hat{f}_{T,i} - \boldsymbol{J}^{-1}\boldsymbol{\omega}_p^{\times}\boldsymbol{J}\boldsymbol{\omega}_p.
\end{aligned} 
\end{equation}
Moreover, from the geometry of the system, the cable vector and its velocity and acceleration have the following relationship:
\begin{equation}\label{eq: length_accel}
\begin{aligned}
& \boldsymbol{l}_j = \boldsymbol{x}_j - \boldsymbol{x}_p -\boldsymbol{R}_{IP}\boldsymbol{t}_j\\
& \dot{\boldsymbol{l}}_j = \boldsymbol{v}_{j, I} - \boldsymbol{v}_{p} + \boldsymbol{R}_{IP}\boldsymbol{t}_j^{\times}\boldsymbol{\omega}_p\\
& \ddot{\boldsymbol{l}}_j  = \dot{\boldsymbol{v}}_{j, I} - \dot{\boldsymbol{v}}_{p} + \boldsymbol{R}_{IP}\boldsymbol{t}_j^{\times}\dot{\boldsymbol{\omega}}_p + \boldsymbol{R}_{IP}\boldsymbol{\omega}_p^{\times}\boldsymbol{t}_j^{\times}\boldsymbol{\omega}_p.
\end{aligned}    
\end{equation}
Hence, combining Eq. \eqref{eq: cable_kinematics}, \eqref{eq: inertial_accel}, \eqref{eq: payload_accel} and \eqref{eq: length_accel}, we have the following:

\begin{equation}\label{eq: cable_dynamics}
\begin{aligned}
&  \ddot{l}_j\boldsymbol{n}_j + l_j \boldsymbol{B}_j\dot{\boldsymbol{v}}_j + 2 \dot{l}_j\boldsymbol{B}_j\boldsymbol{v}_j + l_j\dot{\boldsymbol{B}}_j\boldsymbol{v}_j  =  \frac{1}{m_j} (\boldsymbol{f}_{L, j} -\boldsymbol{n}_j f_{T,j} ) -  \frac{1}{m_p} \bigg(\sum_{i=1}^{N} \boldsymbol{n}_i f_{T,i}\bigg)  \\
& + \boldsymbol{R}_{IP}\boldsymbol{t}_j^{\times}\boldsymbol{J}^{-1}\bigg(\sum_{i=1}^{N} \boldsymbol{t}_i^{\times}\boldsymbol{R}_{PI}\boldsymbol{n}_i f_{T,i} - \boldsymbol{\omega}_p^{\times}\boldsymbol{J}\boldsymbol{\omega}_p\bigg) + \boldsymbol{R}_{IP}\boldsymbol{\omega}_p^{\times}\boldsymbol{t}_j^{\times}\boldsymbol{\omega}_p\\
& =  \frac{\boldsymbol{f}_{L, j}}{m_j}  -\boldsymbol{n}_j \hat{f}_{T,j}  - \sum_{i=1}^{N} c_i \boldsymbol{n}_i  \hat{f}_{T,i}  + \boldsymbol{R}_{IP}\boldsymbol{t}_j^{\times}\sum_{i=1}^{N}\boldsymbol{Y}_i \boldsymbol{t}_i^{\times}\boldsymbol{R}_{PI}\boldsymbol{n}_i \hat{f}_{T,i} - \boldsymbol{R}_{IP}\boldsymbol{t}_j^{\times}\boldsymbol{J}^{-1} \boldsymbol{\omega}_p^{\times}\boldsymbol{J}\boldsymbol{\omega}_p + \boldsymbol{R}_{IP}\boldsymbol{\omega}_p^{\times}\boldsymbol{t}_j^{\times}\boldsymbol{\omega}_p.
\end{aligned}   
\end{equation}

The acceleration of the cable vector $\boldsymbol{l}_j$ can be separated into two channels using the properties in Lemma \ref{lem: B_property}. Multiplying $\boldsymbol{n}_j^T$ to both sides of the Eq. \eqref{eq: cable_dynamics} gives us the dynamics in the channel of cable length $l_j$, such that
\begin{equation}\label{eq: l_dynamics}
    \ddot{\boldsymbol{l}} = \hat{\boldsymbol{f}}_{||} + \boldsymbol{A}_{l} \hat{\boldsymbol{f}}_T +\boldsymbol{\eta} 
\end{equation}
where $\boldsymbol{l} = [l_1, \dots, l_N]^T \in \mathbb{R}^{N \times 1}$, $\boldsymbol{\hat{f}}_{||} = [\hat{f}_{||,1}, \dots, \hat{f}_{||,N}]^T \in \mathbb{R}^{N \times 1}$ and $\boldsymbol{\hat{f}}_T= [\hat{f}_{T,1}, \dots, \hat{f}_{T,N}]^T \in \mathbb{R}^{N \times 1}$ are obtained by stacking variables. $\boldsymbol{A}_l$ and $\boldsymbol{\eta}$ are defined as
\begin{equation}\label{eq: l_ddot_matrices}
\begin{aligned}
\boldsymbol{A}_{l}  &= - \boldsymbol{I}_N + \begin{bmatrix}
     - c_1\boldsymbol{n}_1^T\boldsymbol{n}_1, & \dots &   - c_N \boldsymbol{n}_1^T\boldsymbol{n}_N\\
     \vdots &  \ddots & \vdots  \\
        - c_1 \boldsymbol{n}_N^T\boldsymbol{n}_1, & \dots &   - c_N \boldsymbol{n}_N^T\boldsymbol{n}_N\\
\end{bmatrix} +  \begin{bmatrix}
\boldsymbol{n}_1^T\boldsymbol{R}_{IP}\boldsymbol{t}_1^{\times}\boldsymbol{Y}_1\boldsymbol{t}_1^{\times}\boldsymbol{R}_{PI}\boldsymbol{n}_1, & \dots &  \boldsymbol{n}_1^T\boldsymbol{R}_{IP}\boldsymbol{t}_1^{\times}\boldsymbol{Y}_N\boldsymbol{t}_N^{\times}\boldsymbol{R}_{PI}\boldsymbol{n}_N\\
 \vdots &  \ddots & \vdots\\
\boldsymbol{n}_N^T\boldsymbol{R}_{IP}\boldsymbol{t}_N^{\times}\boldsymbol{Y}_1\boldsymbol{t}_1^{\times}\boldsymbol{R}_{PI}\boldsymbol{n}_1, & \dots &  \boldsymbol{n}_N^T\boldsymbol{R}_{IP}\boldsymbol{t}_N^{\times}\boldsymbol{Y}_N\boldsymbol{t}_N^{\times}\boldsymbol{R}_{PI}\boldsymbol{n}_N
\end{bmatrix}\\
\boldsymbol{\eta} &= \begin{bmatrix}
      - \boldsymbol{n}_1^T  l_1\dot{\boldsymbol{B}}_1\boldsymbol{v}_1 -   \boldsymbol{n}_1^T \boldsymbol{R}_{IP}\boldsymbol{t}_1^{\times}\boldsymbol{J}^{-1}\boldsymbol{\omega}_p^{\times}\boldsymbol{J}\boldsymbol{\omega}_p + \boldsymbol{n}_1^T\boldsymbol{R}_{IP}\boldsymbol{\omega}_p^{\times}\boldsymbol{t}_1^{\times}\boldsymbol{\omega}_p \\ 
      \vdots \\
       - \boldsymbol{n}_N^T  l_N\dot{\boldsymbol{B}}_N\boldsymbol{v}_N -   \boldsymbol{n}_N^T \boldsymbol{R}_{IP}\boldsymbol{t}_N^{\times}\boldsymbol{J}^{-1}\boldsymbol{\omega}_p^{\times}\boldsymbol{J}\boldsymbol{\omega}_p + \boldsymbol{n}_N^T\boldsymbol{R}_{IP}\boldsymbol{\omega}_p^{\times}\boldsymbol{t}_N^{\times}\boldsymbol{\omega}_p 
    
\end{bmatrix}.
\end{aligned}    
\end{equation}
Moreover, multiplying $\boldsymbol{B}_j^T$ to both sides of the Eq. \eqref{eq: cable_dynamics} gives us the dynamics in the channel of $\boldsymbol{v}_j$, such that
\begin{equation}\label{eq: v_dynamics}
\dot{\boldsymbol{v}} =  \boldsymbol{z} + \boldsymbol{A}_v\hat{\boldsymbol{f}}_T + \boldsymbol{\mu},   
\end{equation}
where $\boldsymbol{v} = [\boldsymbol{v}_1^T,\dots,\boldsymbol{v}_N^T]^T\in \mathbb{R}^{2N \times 1}$ and $\boldsymbol{z} = [\boldsymbol{z}_1^T, \dots, \boldsymbol{z}_N^T]^T \in \mathbb{R}^{2N \times 1}$ are obtained by stacking variables. $\boldsymbol{A}_v$ and $\boldsymbol{\mu}$ are defined as 
\begin{equation}\label{eq: v_dot_matrices}
\begin{aligned}
\boldsymbol{A}_v &= \begin{bmatrix}
    -   \boldsymbol{\Xi}_1  c_1\boldsymbol{n}_1 & \dots, &   -   \boldsymbol{\Xi}_1  c_N\boldsymbol{n}_N \\
    \vdots & \ddots & \vdots\\
    -   \boldsymbol{\Xi}_N  c_1\boldsymbol{n}_1 & \dots, &   -   \boldsymbol{\Xi}_N  c_N\boldsymbol{n}_N 
\end{bmatrix} + \begin{bmatrix}
\boldsymbol{\Xi}_1\boldsymbol{R}_{IP}\boldsymbol{t}_1^{\times}\boldsymbol{Y}_1\boldsymbol{t}_1^{\times}\boldsymbol{R}_{PI}\boldsymbol{n}_1, & \dots &, \boldsymbol{\Xi}_1\boldsymbol{R}_{IP}\boldsymbol{t}_1^{\times}\boldsymbol{Y}_N\boldsymbol{t}_N^{\times}\boldsymbol{R}_{PI}\boldsymbol{n}_N\\
\vdots & \ddots & \vdots\\
\boldsymbol{\Xi}_N\boldsymbol{R}_{IP}\boldsymbol{t}_N^{\times}\boldsymbol{Y}_1\boldsymbol{t}_1^{\times}\boldsymbol{R}_{PI}\boldsymbol{n}_1, & \dots &, \boldsymbol{\Xi}_N\boldsymbol{R}_{IP}\boldsymbol{t}_N^{\times}\boldsymbol{Y}_N\boldsymbol{t}_N^{\times}\boldsymbol{R}_{PI}\boldsymbol{n}_N
\end{bmatrix},\\
\boldsymbol{\mu} &= \begin{bmatrix}
    -  \frac{2\dot{l}_1}{l_1} \boldsymbol{v}_1 - \boldsymbol{\Xi}_1l_1\dot{\boldsymbol{B}}_1\boldsymbol{v}_1  - \boldsymbol{\Xi}_1  \boldsymbol{R}_{IP}\boldsymbol{t}_1^{\times}\boldsymbol{J}^{-1} \boldsymbol{\omega}_p^{\times}\boldsymbol{J}\boldsymbol{\omega}_p + \boldsymbol{\Xi}_1 \boldsymbol{R}_{IP}\boldsymbol{\omega}_p^{\times}\boldsymbol{t}_1^{\times}\boldsymbol{\omega}_p \\
    \vdots \\
    -  \frac{2\dot{l}_N}{l_N} \boldsymbol{v}_N - \boldsymbol{\Xi}_Nl_N\dot{\boldsymbol{B}}_N\boldsymbol{v}_N  - \boldsymbol{\Xi}_N  \boldsymbol{R}_{IP}\boldsymbol{t}_N^{\times}\boldsymbol{J}^{-1} \boldsymbol{\omega}_p^{\times}\boldsymbol{J}\boldsymbol{\omega}_p + \boldsymbol{\Xi}_N \boldsymbol{R}_{IP}\boldsymbol{\omega}_p^{\times}\boldsymbol{t}_N^{\times}\boldsymbol{\omega}_p
\end{bmatrix}.
\end{aligned}    
\end{equation}
Rewriting the dynamics of the payload in Eq. \eqref{eq: payload_accel} into a system of equations gives us
\begin{equation}\label{eq: payload_wrench}
\begin{bmatrix}
         \dot{\boldsymbol{v}}_{p}\\
      \dot{\boldsymbol{\omega}}_p 
\end{bmatrix} =  \boldsymbol{A}_p \hat{\boldsymbol{f}}_T + \boldsymbol{\gamma},
\end{equation}
where $\boldsymbol{A}_p = \begin{bmatrix}
      c_1\boldsymbol{n}_1 & \dots & c_N\boldsymbol{n}_N\\
        \boldsymbol{Y}_1\boldsymbol{t}_1^{\times}\boldsymbol{R}_{PI}\boldsymbol{n}_1 &  \dots & \boldsymbol{Y}_N\boldsymbol{t}_N^{\times}\boldsymbol{R}_{PI}\boldsymbol{n}_N 
\end{bmatrix} \in \mathbb R^{6\times N}$ and $\boldsymbol{\gamma} =  \begin{bmatrix}
    \boldsymbol{g}_I\\
     - \boldsymbol{J}^{-1} \boldsymbol{\omega}_p^{\times}\boldsymbol{J}\boldsymbol{\omega}_p
\end{bmatrix} \in \mathbb R^{6\times 1}$. Defining auxiliary control inputs $\tilde{\boldsymbol{z}} = \boldsymbol{z} + \boldsymbol{A}_v\hat{\boldsymbol{f}}_T +  \boldsymbol{\mu}$, $\tilde{\boldsymbol{f}}_{||} =  \hat{\boldsymbol{f}}_{||} + \boldsymbol{A}_{l} \hat{\boldsymbol{f}}_T  +  \boldsymbol{\eta}$, the following equations of motion can be obtained:
\begin{equation}\label{eq: variable_length_control_affine_payload}
\begin{aligned}
& \Sigma_p: \quad
\left\{
\begin{aligned}
& \begin{bmatrix}
    \dot{\boldsymbol{v}}\\
    \dot{\boldsymbol{v}}_{p}\\
    \dot{\boldsymbol{\omega}}_p\\
    \dot{\boldsymbol{r}}\\
    \dot{\boldsymbol{x}}_p
\end{bmatrix} =  \begin{bmatrix}
      \boldsymbol{0}_{2N \times 1}\\
      \boldsymbol{\gamma}\\
      \boldsymbol{v}\\
      \boldsymbol{v}_p
\end{bmatrix}  + \begin{bmatrix}
     \boldsymbol{I}_{2N} & \boldsymbol{0}_{2N\times N}\\
      \boldsymbol{0}_{6\times 2N} & \boldsymbol{A}_p \\
       \boldsymbol{0}_{2N\times 2N} & \boldsymbol{0}_{2N\times N}  \\
       \boldsymbol{0}_{3\times 2N} & \boldsymbol{0}_{3\times N}  \\
\end{bmatrix} \begin{bmatrix}
    \tilde{\boldsymbol{z}} \\
     \hat{\boldsymbol{f}}_T 
\end{bmatrix} \\
& \dot{\boldsymbol{R}}_{IP} = \boldsymbol{R}_{IP}\boldsymbol{\omega}_p^{\times}
\end{aligned}
\right.
\\
\end{aligned} 
\end{equation}

\begin{equation}\label{eq: variable_length_control_affine_length}
\begin{aligned}
& \Sigma_l: \quad
\left\{
\begin{aligned}
& \begin{bmatrix}
    \dot{\boldsymbol{l}}\\
    \ddot{\boldsymbol{l}}
\end{bmatrix}
=
\begin{bmatrix}
    \dot{\boldsymbol{l}}\\
    \boldsymbol{0}_{N \times 1}
\end{bmatrix}
+
\begin{bmatrix}
    \boldsymbol{0}_{N \times N}\\
    \boldsymbol{I}_{N}
\end{bmatrix}
\tilde{\boldsymbol{f}}_{||}
\end{aligned}
\right.
\end{aligned}
\end{equation}

From this formulation, the system $\Sigma_p$ is decoupled from the system $\Sigma_l$. Lastly, as the cables are attached at the C.M. of the quadrotors from Assumption~\ref{assum: system}, the decoupled attitude dynamics of the $j^{th}$ quadrotor can be obtained as follows:
\begin{equation}\label{eq: variable_length_control_affine_attitude}
\Sigma_j: \quad  \quad
\left\{
\begin{aligned}
&\dot{\boldsymbol{\omega}}_j = \boldsymbol{J}^{-1}_j(\boldsymbol{\tau}_j - \boldsymbol{\omega}_j^{\times}\boldsymbol{J}_j\boldsymbol{\omega}_j)\\
& \dot{\boldsymbol{R}}_{Ij} = \boldsymbol{R}_{Ij}\boldsymbol{\omega}_j^{\times}
\end{aligned}
\right.
\end{equation}

\section{Controller Design and Implementation}\label{sec: controller_design_and_implementation}

\subsection{Proposed Control Framework for the Complete System}

From the system modelling in Eq.~\eqref{eq: variable_length_control_affine_payload},~\eqref{eq: variable_length_control_affine_length},~\eqref{eq: variable_length_control_affine_attitude}, the overall dynamics are decomposed into the payload subsystem $\Sigma_p$ and the cable-length subsystem $\Sigma_l$, and the quadrotor attitude subsystem $\Sigma_j$. Since the cable length states do not explicitly appear in $\Sigma_p$, the system admits a decoupled control structure, enabling independent controller design for payload manipulation and cable length regulation. 

\begin{figure}[htbp]
    \centering
    \includegraphics[width=0.95\textwidth]{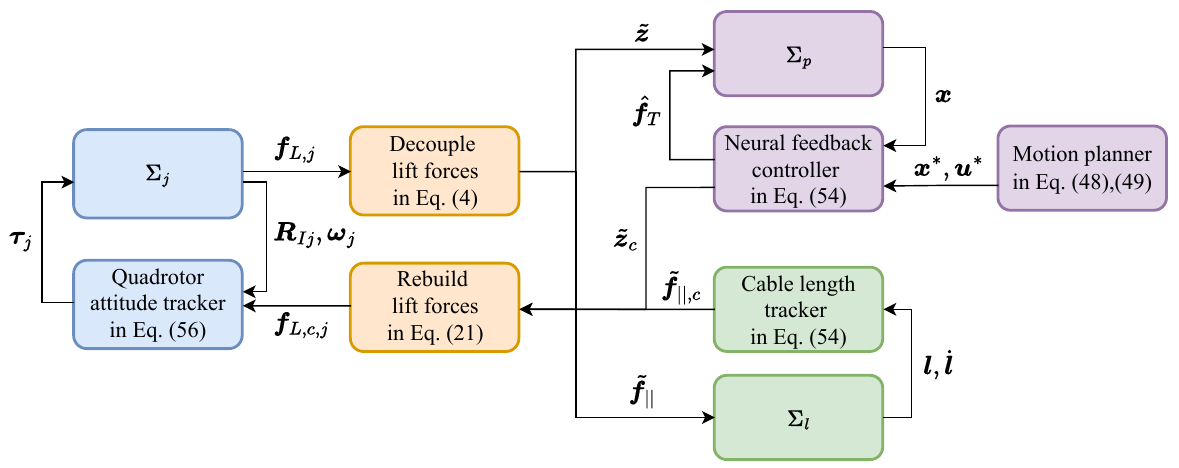}
    \caption{Control framework of the complete system.}
    \label{fig: control_framework}
\end{figure}

Therefore, the control framework is proposed in Fig.~\ref{fig: control_framework}, where the payload dynamics and cable length dynamics are controlled independently while remaining physically coupled through the generated lift forces. For the payload subsystem $\Sigma_p$, a neural CCM-based controller is designed to regulate the payload movement and cable direction dynamics. In particular, the controller generates the desired cable-direction acceleration input $\tilde{\boldsymbol{z}}_c$ and the required normalized cable tension $\hat{\boldsymbol{f}}_{T,c}$ for each quadrotor.

Separately, the cable-length subsystem $\Sigma_l$ is controlled to track a desired cable length trajectory using an independent length controller, generating $\tilde{\boldsymbol{f}}_{||,c}$. The cable-length controller enables active reconfiguration of the multi-drone formation during flight, allowing the system to exploit the additional degrees of freedom introduced by the variable-length cables.

The outputs of the payload and cable-length controllers $\tilde{\boldsymbol{z}}_c,\hat{\boldsymbol{f}}_{T,c},\tilde{\boldsymbol{f}}_{||,c}$ are combined to recover the desired lift force vector for each quadrotor. $\boldsymbol{z}_{c,j}$ and $\hat{f}_{||,c,j}$ can be back-computed by the definitions of auxiliary control , given the current states and $\hat{\boldsymbol{f}}_{T,c}$, such that
\begin{equation}\label{eq: auxiliary_control_definitions}
    \boldsymbol{z}_{c} = \tilde{\boldsymbol{z}}_c - \boldsymbol{A}_v\hat{\boldsymbol{f}}_{T,c} -  \boldsymbol{\mu}, \quad \hat{\boldsymbol{f}}_{||,c} = \tilde{\boldsymbol{f}}_{||,c} - \boldsymbol{A}_{l} \hat{\boldsymbol{f}}_{T,c} -  \boldsymbol{\eta}.
\end{equation}
The desired lift vector $\boldsymbol{f}_{L,c,j}$ for the $j^{th}$ can then be recovered as
\begin{equation}\label{eq: recover_lift}
    \boldsymbol{f}_{L,c,j} = m_j \left(l_j\boldsymbol{B}_j\boldsymbol{z}_{c,j} + \boldsymbol{n}_j \hat{f}_{||,c,j}\right).
\end{equation}

Hence, $\boldsymbol{f}_{L,c,j}$ is subsequently tracked by the individual quadrotors through an inner-loop geometric controller as the attitude dynamics $\Sigma_j$ are decoupled, ensuring that each drone produces the required lift vector. This cascaded control structure allows for independent controller design for each subsystem. 

\subsection{Neural Contraction Control for Payload Subsystem}\label{sec: training_payload_subsystem}

The control-affine system in $\Sigma_p$ is underactuated and highly nonlinear. Therefore, a learning framework is adopted to train a contracting neural controller. Building on the work of \cite{Lo2026}, which establishes a framework for jointly training a Control Contraction Metric (CCM) and a neural controller for a body-rate controlled quadrotor over the state space $\mathbb R^6\times SO(3)$, analogous methods are employed here to synthesize a feedback controller for $\Sigma_p$. Specifically, a three-drone configuration ($N$=3) is considered, corresponding to the state space $\mathcal{X}=\mathbb{R}^{21}\times SO(3)$. While the proposed framework naturally extends to configurations with more drones ($N$>3), a minimum of 3 drones is necessary to achieve full manipulation of both the payload position and attitude.

\subsubsection{The Differential System Dynamics}

The state vector is first defined as $\boldsymbol{x}=[ \boldsymbol{v},\boldsymbol{v}_{p},\boldsymbol{\omega}_p,\boldsymbol{r},\boldsymbol{x}_p,\boldsymbol{r}_{IP}]\in\mathcal{X}$, where $\boldsymbol{r}_{IP}=\mathrm{vec}(\boldsymbol{R}_{IP})$. Then, rewrite Eq. \eqref{eq: variable_length_control_affine_payload} into the the control-affine form $\dot{\boldsymbol{x}}=\boldsymbol{f}(\boldsymbol{x})+\boldsymbol{B}(\boldsymbol{x})\boldsymbol{u}$, where 
\begin{equation}
    \boldsymbol{f}(\boldsymbol{x})=\begin{bmatrix}
      \boldsymbol{0}_{6 \times 1}\\
      \boldsymbol{\gamma}\\
      \boldsymbol{v}\\
      \boldsymbol{v}_p\\
      \boldsymbol{S}_r\boldsymbol{\omega}_p
\end{bmatrix}\in\mathbb{R}^{30} ,\quad \boldsymbol{B}(\boldsymbol{x})=\begin{bmatrix}
     \boldsymbol{I}_{6} & \boldsymbol{0}_{6\times 3}\\
      \boldsymbol{0}_{6\times 6} & \boldsymbol{A}_p \\
       \boldsymbol{0}_{6\times 6} & \boldsymbol{0}_{6\times 3}  \\
       \boldsymbol{0}_{3\times 6} & \boldsymbol{0}_{3\times 3}  \\
       \boldsymbol{0}_{9\times 6} & \boldsymbol{0}_{9\times 3}
\end{bmatrix}\in\mathbb{R}^{30\times9}, \quad\boldsymbol{u}=\begin{bmatrix}
    \tilde{\boldsymbol{z}} \\
     \hat{\boldsymbol{f}}_T 
\end{bmatrix}\in\mathbb{R}^{9}
\end{equation}
Note that $\boldsymbol{S}_r = [\mathrm{vec}(\boldsymbol{R}_{IP}\boldsymbol{e}_1^\times), \,\mathrm{vec}(\boldsymbol{R}_{IP}\boldsymbol{e}_2^\times), \,\mathrm{vec}(\boldsymbol{R}_{IP}\boldsymbol{e}_3^\times)]\in\mathbb R^{9\times 3}$, and $\boldsymbol{A}_p\in\mathbb{R}^{6\times 3}$ according to the definitions in Eq.~\eqref{eq: payload_wrench}. Defining ($\boldsymbol{x}^*$,$\boldsymbol{u}^*$) as the reference trajectory from motion planning, a feedback control law is formulated as
\begin{equation}\label{eq: fb_ctrl_format}
    \boldsymbol{u} = \boldsymbol{k}(\boldsymbol{x}, \boldsymbol{x}^*) + \boldsymbol{u}^*, 
\end{equation}
where $\boldsymbol{k}(\boldsymbol{x}, \boldsymbol{x}^*)=\boldsymbol{0}$ such that the pair ($\boldsymbol{x}^*$,$\boldsymbol{u}^*$) is a particular solution to the closed-loop system. 

A standard procedure for designing a contraction controller is to analyze the differential system \cite{Manchester2017a}. The differential system associated with the system in $\Sigma_p$ can be easily obtained as
\begin{equation}\label{eq: differential_system_payload}
    \delta\dot{\boldsymbol{x}} = \boldsymbol{A}(\boldsymbol{x}, \boldsymbol{u})\delta\boldsymbol{x} + \boldsymbol{B}(\boldsymbol{x})\delta\boldsymbol{u},
\end{equation}
where $\boldsymbol{A}(\boldsymbol{x}, \boldsymbol{u}) = \frac{\partial \boldsymbol{f}}{\partial \boldsymbol{x}} + \sum_{i=1}^m \frac{\partial \boldsymbol{b}_i}{\partial \boldsymbol{x}} u_{i}$. Note that $\boldsymbol{b}_i$ represents the $i^{th}$ column of $\boldsymbol{B}(\boldsymbol{x})$ and $u_i$ represents the $i^{th}$ element of $\boldsymbol{u}$. With the feedback controller in Eq. \eqref{eq: fb_ctrl_format}, the closed-loop differential system can be written as 
\begin{equation}\label{differential_closed_loop_system_payload}
    \delta\dot{\boldsymbol{x}} = \bigg(\boldsymbol{A}(\boldsymbol{x}, \boldsymbol{u}) + \boldsymbol{B}(\boldsymbol{x})\boldsymbol{K}(\boldsymbol{x})\bigg)\delta\boldsymbol{x},
\end{equation}
where $\boldsymbol{K}(\boldsymbol{x})=\frac{\partial\boldsymbol{k}}{\partial\boldsymbol{x}}$. 

For systems evolving on Lie groups, the differential states $\delta\boldsymbol{x}$ and vector fields $\boldsymbol{f}(\boldsymbol{x})$ and $\boldsymbol{\boldsymbol{B}}(\boldsymbol{x})$ should satisfy transversality conditions \cite{Wu2014}, where the associated vectors are always tangential to the manifold. For Lie group state space $\mathcal{X}$, it is a 24-dimensional manifold embedded in a 30-dimensional Euclidean space. Therefore, there exist global vector fields $\{\boldsymbol{s}_1,\, \dots, \boldsymbol{s}_{24}\}$ that form a basis of the tangent space $T_{\boldsymbol{x}}\mathcal{X}$ at each $\boldsymbol{x}\in\mathcal{X}$. Subsequently, a smooth function $\boldsymbol{S}(\boldsymbol{x}):=\begin{bmatrix}
    \boldsymbol{s}_1,\,\dots, \boldsymbol{s}_{24}
\end{bmatrix} \in \mathbb R^{30\times 24}$ can be defined such that 
\begin{equation}\label{eq: extrinsic_to_intrinsic}
    \forall\boldsymbol{x}\in\mathcal X,\,\,T_{\boldsymbol{x}}\mathcal X=\{ \boldsymbol{S}(\boldsymbol{x})\boldsymbol{\vartheta}\mid\boldsymbol{\vartheta}\in \mathbb{R}^{24}\}, 
\end{equation}
where $\boldsymbol{\vartheta}$ is the intrinsic representation of tangent vectors. In particular, it can be easily shown that the tangent space of $\mathbb R^n$ is spanned by Euclidean basis vectors and the tangent space of $SO(3)$ is spanned by the columns of $\boldsymbol{S}_r$. Hence, the $\boldsymbol{S}(\boldsymbol{x})$ matrix for $\Sigma_p$ is chosen as 
\begin{equation}
    \begin{bmatrix}
        \boldsymbol{I}_{21} & \boldsymbol{0}_{21\times 3} \\
        \boldsymbol{0}_{9\times 21} & \frac{1}{\sqrt{2}}\boldsymbol{S}_r
    \end{bmatrix},
\end{equation}
where $\boldsymbol{S}^\top\boldsymbol{S}=\boldsymbol{I}_{24}$ as $\boldsymbol{S}_r^\top\boldsymbol{S}_r=2\boldsymbol{I}_3$. Smooth functions $\boldsymbol{\vartheta}\in \mathbb R^{24}$ and $\boldsymbol{E}(\boldsymbol{x})\in \mathbb R^{24\times 9}$ can always be found such that 
\begin{equation}\label{eq: E_matrix_transformation}
    \delta\boldsymbol{x} = \boldsymbol{S}\boldsymbol{\vartheta},\quad\boldsymbol{B}=\boldsymbol{S}\boldsymbol{E}.
\end{equation}
The intrinsic representation of the differential system can then be analysed. Given the feedback controller in the format in Eq. \eqref{eq: fb_ctrl_format} and defining $\boldsymbol{P}_S=\boldsymbol{S}(\boldsymbol{S}^\top\boldsymbol{S})^{-1}$, 
\begin{equation}
    \dot{\boldsymbol{\vartheta}}=\frac{d}{dt}(\boldsymbol{P}_S^\top\delta\boldsymbol{x}) = \dot{\boldsymbol{P}}_S^\top\boldsymbol{S}\boldsymbol{\vartheta} + \boldsymbol{P}_S^\top\delta\dot{\boldsymbol{x}} = (\dot{\boldsymbol{P}}_S^\top+\boldsymbol{P}_S^\top\boldsymbol{A}+\boldsymbol{E}\boldsymbol{K})\boldsymbol{S}\boldsymbol{\vartheta} = \boldsymbol{\mathcal{A}}\boldsymbol{\vartheta}, 
\end{equation}
where $\boldsymbol{\mathcal{A}}=(\dot{\boldsymbol{P}}_S^\top+\boldsymbol{P}_S^\top\boldsymbol{A}+\boldsymbol{E}\boldsymbol{K})\boldsymbol{S}$ and $\dot{\boldsymbol{P}}_S^\top=\partial_{\dot{\boldsymbol{x}}}\boldsymbol{P}_S^\top$. Note that for our system, $\boldsymbol{P}_S=\boldsymbol{S}\cdot\boldsymbol{I}_{24} = \boldsymbol{S}$.

\subsubsection{Control Contraction Metrics on Lie Groups}

With the differential system modelled, the feedback controller is now designed using the CCM approach for Lie groups. This involves searching for a feedback controller in the form of Eq. \eqref{eq: fb_ctrl_format} and a Riemannian metric $\boldsymbol{\mathcal{M}}(\boldsymbol{x}):\mathcal{X}\to\mathbb{R}^{24\times24}$ such that $\boldsymbol{\mathcal{M}}(\boldsymbol{x})$ is uniformly bounded (i.e., $\underline{\mathfrak{m}}\boldsymbol{I}_{24}\preceq\boldsymbol{\mathcal{M}}(\boldsymbol{x})\preceq\overline{\mathfrak{m}}\boldsymbol{I}_{24}$ for $\underline{\mathfrak{m}},\overline{\mathfrak{m}}>0$), and the following contraction condition holds for all $\boldsymbol{x}\in\mathcal{X}$
\begin{equation}\label{eq: contraction_condition}
    \boldsymbol{C}_{CCM} = \dot{\boldsymbol{\mathcal{M}}} + \langle\boldsymbol{\mathcal{M}}\boldsymbol{\mathcal{A}} \rangle + 2\lambda \boldsymbol{\mathcal{M}} \preceq \boldsymbol{0},
\end{equation}
where $\lambda>0$ is the contraction rate and $\dot{\boldsymbol{\mathcal{M}}}=\partial_{\dot{\boldsymbol{x}}}\boldsymbol{\mathcal{M}}$. 

Stronger auxiliary conditions are also provided that are useful in learning the contraction controller. Condition C1 ensures the uncontrolled system is contracting in directions orthogonal to the span of the control inputs under metric $\boldsymbol{\mathcal{M}}$, 
\begin{equation}\label{eq: C1}
    \boldsymbol{C}_1=\boldsymbol{E}_{\perp}^\top\left( -\partial_{\boldsymbol{f}}\boldsymbol{\mathcal W} + \langle\boldsymbol{S}_{\boldsymbol{f}}\boldsymbol{\mathcal W}\rangle + 2\lambda\boldsymbol{\mathcal W}\right)\boldsymbol{E}_{\perp} \prec \boldsymbol{0},
\end{equation}
where $\boldsymbol{E}_{\perp}$ is any annihaltor matrix of $\boldsymbol{E}$ such that $\boldsymbol{E}_{\perp}^\top\boldsymbol{E}=\boldsymbol{0}$, $\boldsymbol{\mathcal W}=\boldsymbol{\mathcal M}^{-1}$ and $\boldsymbol{S}_{\boldsymbol{f}}=(\partial_{\boldsymbol{f}}\boldsymbol{P}_S^\top+\boldsymbol{P}_S^\top\frac{\partial \boldsymbol{f}}{\partial\boldsymbol{x}})\boldsymbol{S}$. Condition C2 enforces vector fields $\boldsymbol{b}_i$ to be Killing vector fields, 
\begin{equation}\label{eq: C2}
    \boldsymbol{C}_{2,i}=\boldsymbol{E}_{\perp}^\top\big( -\partial_{\boldsymbol{b}_i}\boldsymbol{\mathcal W} + \langle\boldsymbol{S}_{\boldsymbol{b}_i}\boldsymbol{\mathcal W}\rangle \big)\boldsymbol{E}_{\perp} = \boldsymbol{0}, \quad (i = 1,2,\dots,9),
\end{equation}
where $\boldsymbol{S}_{\boldsymbol{b}_i}=(\partial_{\boldsymbol{b}_i}\boldsymbol{P}_S^\top+\boldsymbol{P}_S^\top\frac{\partial \boldsymbol{b}_i}{\partial\boldsymbol{x}})\boldsymbol{S}$. The two conditions together guarantee the existence of a CCM $\boldsymbol{\mathcal{M}}$ and a feedback controller in the form of Eq. \eqref{eq: fb_ctrl_format} such that the closed-loop system is contracting with rate $\lambda$ \cite{Wu2024}. Note that by inspection, $\boldsymbol{E}(\boldsymbol{x})$ and $\boldsymbol{E}_\perp(\boldsymbol{x})$ can be obtained as
\begin{equation}
    \boldsymbol{E}(\boldsymbol{x})=\begin{bmatrix}
     \boldsymbol{I}_{6} & \boldsymbol{0}_{6\times 3}\\
      \boldsymbol{0}_{6\times 6} & \boldsymbol{A}_p \\
       \boldsymbol{0}_{6\times 6} & \boldsymbol{0}_{6\times 3}  \\
       \boldsymbol{0}_{3\times 6} & \boldsymbol{0}_{3\times 3}  \\
       \boldsymbol{0}_{3\times 6} & \boldsymbol{0}_{3\times 3}
\end{bmatrix}, \quad \boldsymbol{E}_\perp(\boldsymbol{x}) = \begin{bmatrix}
     \boldsymbol{0}_{6\times3} & \boldsymbol{0}_{6\times 6} & \boldsymbol{0}_{6\times3} & \boldsymbol{0}_{6\times3}\\
      \boldsymbol{A}_{p,\perp} & \boldsymbol{0}_{3\times 6}  & \boldsymbol{0}_{3\times3} & \boldsymbol{0}_{3\times3}\\
       \boldsymbol{0}_{6\times 3} & \boldsymbol{I}_{6} & \boldsymbol{0}_{6\times3} & \boldsymbol{0}_{6\times3}\\
       \boldsymbol{0}_{3\times 3} & \boldsymbol{0}_{3\times 6}  & \boldsymbol{I}_{3} & \boldsymbol{0}_{3\times3} \\
       \boldsymbol{0}_{3\times 3} & \boldsymbol{0}_{3\times 6} & \boldsymbol{0}_{3\times3} & \boldsymbol{I}_{3}
\end{bmatrix}\in\mathbb R^{24\times 15},
\end{equation}
where $\boldsymbol{A}_{p,\perp}\in\mathbb R^{3\times 3}$ spans the null space of $\boldsymbol{A}_p$ and can be computed numerically using singular value decomposition.

\subsubsection{Training Framework}

\begin{figure}[htbp]
    \centering
    \includegraphics[width=0.7\textwidth]{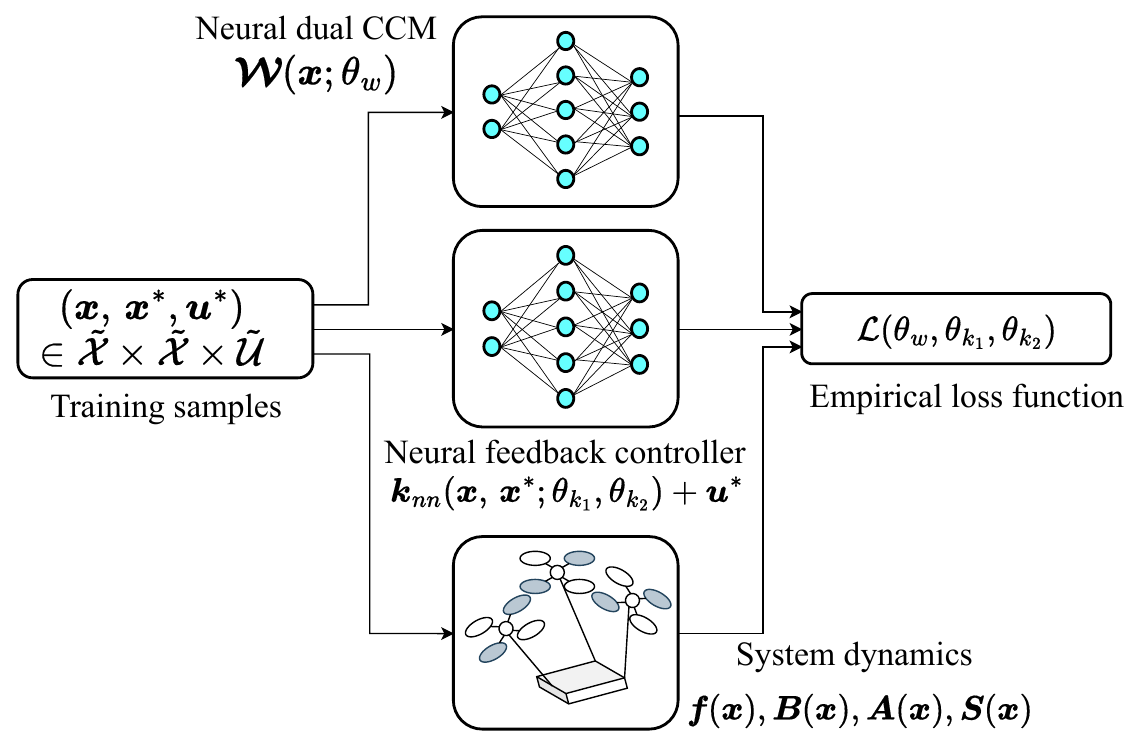}
    \caption{Training framework for dual synthesis of neural CCM and neural feedback controller.}
    \label{fig: training_framework}
\end{figure}

Following the training framework proposed in \cite{Sun2021,Lo2026}, both the CCM and neural control are modelled as neural networks and trained simultaneously as illustrated in Fig.~\ref{fig: training_framework}. The neural feedback controller is modelled as $\boldsymbol{u}=\boldsymbol{k}_{nn}(\boldsymbol{x},\boldsymbol{x}^*;\theta_{k_1},\theta_{k_2})+\boldsymbol{u}^*$, where $\boldsymbol{k}_{nn}(\boldsymbol{x},\boldsymbol{x}^*;\theta_{k_1},\theta_{k_2})=\boldsymbol{K}_1(\boldsymbol{x},\boldsymbol{x}^*;\theta_{k_1})\cdot \tanh{\big(\boldsymbol{K}_2(\boldsymbol{x},\boldsymbol{x}^*;\theta_{k_2})\cdot\boldsymbol{\varepsilon}(\boldsymbol{x},\boldsymbol{x}^*)}\big)$. $\theta_{k_1}$ and $\theta_{k_2}$ are neural network weights for $\boldsymbol{K}_1\in\mathbb{R}^{9\times90}$ and $\boldsymbol{K}_2\in\mathbb{R}^{90\times 24}$, and $\tanh{(\cdot)}$ is the hyperbolic tangent function. $\boldsymbol{\varepsilon}(\cdot,\cdot):\mathcal{X}\times\mathcal{X}\to\mathbb R^{24}$ is an error function for Lie groups represented in Euclidean space with $\boldsymbol{\varepsilon}(\boldsymbol{x},\boldsymbol{x})=\boldsymbol{0}$ $\forall \boldsymbol{x}\in\mathcal{X}$. For our system, the error in Lie algebra is used for states in $SO(3)$ following \cite{Lo2026}, such that 
\begin{equation}
    \boldsymbol{\varepsilon}(\boldsymbol{x},\boldsymbol{x}^*)=\begin{bmatrix}
        \boldsymbol{\varepsilon}_e^\top & \boldsymbol{\varepsilon}_r^\top
    \end{bmatrix}^\top, \quad \boldsymbol{\varepsilon}_e=[\boldsymbol{I}_{21}\,\,\boldsymbol{0}_{21\times9}](\boldsymbol{x}-\boldsymbol{x}^*)\in\mathbb{R}^{21}, \quad \boldsymbol{\varepsilon}_r=\frac{1}{2}(\boldsymbol{R}_{IB}^{*\top}\boldsymbol{R}_{IB}-\boldsymbol{R}_{IB}^\top\boldsymbol{R}_{IB}^*)^\vee\in\mathbb{R}^{3}. 
\end{equation}
This formulation follows the structure of feedback controller modelled in Eq. \eqref{eq: fb_ctrl_format}, where $\boldsymbol{k}_{nn}(\boldsymbol{x},\boldsymbol{x})=\boldsymbol{0}\,\,\forall\boldsymbol{x}\in\mathcal{X}$. Moreover, the dual metric is modelled as $\boldsymbol{\mathcal{W}}(\boldsymbol{x};\theta_w)=\boldsymbol{\Theta}(\boldsymbol{x};\theta_{w})^\top\boldsymbol{\Theta}(\boldsymbol{x};\theta_{w})+\overline{\mathfrak{m}}^{-1}\boldsymbol{I}_q$, such that $\underline{\mathfrak{m}}\boldsymbol{1}\preceq\boldsymbol{\mathcal{M}}$ by construction. 

Let $\{(\boldsymbol{x}_k,\boldsymbol{x}^*_k,\boldsymbol{u}^*_k\}^K_{k=1}$ be $K$ training samples drawn uniformly from $\tilde{\mathcal{X}}\times\tilde{\mathcal{X}}\times\tilde{\mathcal{U}}$, where $\tilde{\mathcal{X}}\subseteq\mathcal{X}$ and $\tilde{\mathcal{U}}\subseteq\mathbb{R}^9$ are compact subsets of the state space and control space respectively. The network parameters $\theta_{k_1}$, $\theta_{k_2}$ and $\theta_{w}$ are trained by minimizing the empirical loss $\mathcal{L}=\frac{1}{K}\sum^K_{k=1}\mathcal{L}_k$, where
\begin{equation}\label{eq: training_loss_k}
    \mathcal{L}_k = \operatorname{\mathcal{L}_{PD}}(-\boldsymbol{C}_{CCM}) + 
    \operatorname{\mathcal{L}_{PD}}(-\boldsymbol{C}_1) + \sum_{i=1}^9\|\boldsymbol{C}_{2,i}\|_F + \operatorname{\mathcal{L}_{PD}}(\underline{\mathfrak{m}}^{-1}\boldsymbol{I}_{24} - \boldsymbol{\mathcal{W}}) 
\end{equation}
is the loss for each training sample. Note that $\|\cdot\|_F$ is the Frobenius norm. Also, given a matrix $\boldsymbol{A}\in\mathbb R^{n\times n}$, $\operatorname{\mathcal{L}_{PD}}(\boldsymbol{A})\geq 0$ penalizes negative definiteness of $\boldsymbol{A}$. Uniformly sampling 1024 unit vectors to get the set $\{\boldsymbol{p}_i\in\mathbb{R}^n \mid \|\boldsymbol{p}_i\|=1 \}^{1024}_{i=1}$, the positive-definiteness loss is computed as  $\operatorname{L_{PD}}(\boldsymbol{A})=\frac{1}{1024}\sum^{1024}_{i=1}\max(0,-\boldsymbol{p}_i^\top \boldsymbol{A}\boldsymbol{p}_i)$. The first three soft loss term in Eq. \eqref{eq: training_loss_k} penalizes the violations of conditions in Eq. \eqref{eq: contraction_condition}, \eqref{eq: C1} and \eqref{eq: C2} respectively. Moreover, the loss term $\operatorname{\mathcal{L}_{PD}}(\underline{\mathfrak{m}}^{-1}\boldsymbol{I}_{24} - \boldsymbol{\mathcal{W}})$ penalizes the learned dual metric if it exceeds the upper bound $\underline{\mathfrak{m}}^{-1}$, enforcing the uniform boundedness of the CCM $\boldsymbol{\mathcal{M}}(\boldsymbol{x})$. The forward pass of the training framework is illustrated in Fig.~\ref{fig: training_framework}, where the training samples and the current network weights are used to compute the neural dual CCM and the neural feedback controller. Together with the system dynamics, the empirical loss function can be obtained by measuring the violations in contracting conditions as in Eq. \eqref{eq: training_loss_k}, and the neural network weights are optimized through gradient descent. 

\subsubsection{Motion Planning via Differential Flatness and Tension Allocation}\label{sec: motion_planning}
A smooth reference trajectory $(\boldsymbol{x}^*, \boldsymbol{u}^*)$ consistent with the system dynamics must be provided to the CCM controller at each time step. Since the payload pose $(\boldsymbol{x}_p, \boldsymbol{R}_{IP})$ and its time derivatives serve as flat outputs of the system, the reference cable directions and tensions can be computed analytically from a desired payload trajectory without numerical integration.

Given a smooth desired payload position trajectory $\boldsymbol{x}_p^*(t)$ and attitude trajectory $\boldsymbol{R}_{IP}^*(t)$, their time derivatives $\boldsymbol{v}_p^*, \dot{\boldsymbol{v}}_p^*$ and $\boldsymbol{\omega}_p^*, \dot{\boldsymbol{\omega}}_p^*$ can be obtained, where $\boldsymbol{\omega}^*_p = (\boldsymbol{R}_{IP}^{*\top}\dot{\boldsymbol{R}}_{IP}^*)^\vee$. The required wrench of the payload $\boldsymbol{w}\in\mathbb{R}^6$ can then be computed following the equations of motion in Eq.~\eqref{eq: payload_accel} as
\begin{equation}\label{eq: wrench_definition}
    \boldsymbol{w}=\begin{bmatrix}
        \boldsymbol{w}_1 \\ \boldsymbol{w}_2
    \end{bmatrix}=\begin{bmatrix}
        \dot{\boldsymbol{v}}^*_p - \boldsymbol{g}_I \\ 
        \dot{\boldsymbol{\omega}}^*_p + \boldsymbol{J}^{-1}\boldsymbol{\omega}_p^{*\times}\boldsymbol{J}\boldsymbol{\omega}^*_p
    \end{bmatrix} = \boldsymbol{A}_p\hat{\boldsymbol{f}}_T.
\end{equation}
For the reference motion generator to also provide the reference control $\boldsymbol{u}^*$, the first and second time derivatives of the wrench are also needed and computed similarly as
\begin{equation}
\begin{aligned}
    \dot{\boldsymbol{w}}_1 &= \ddot{\boldsymbol{v}}^*_p, \quad
    \dot{\boldsymbol{w}}_2 = \ddot{\boldsymbol{\omega}}^*_p + \boldsymbol{J}^{-1}\big(\dot{\boldsymbol{\omega}}_p^{*\times}\boldsymbol{J}\boldsymbol{\omega}^*_p + \boldsymbol{\omega}_p^{*\times}\boldsymbol{J}\dot{\boldsymbol{\omega}}^*_p\big),\\
    \ddot{\boldsymbol{w}}_1 &= \dddot{\boldsymbol{v}}^*_p, \quad
    \ddot{\boldsymbol{w}}_2 = \dddot{\boldsymbol{\omega}}^*_p + \boldsymbol{J}^{-1}\big(\ddot{\boldsymbol{\omega}}_p^{*\times}\boldsymbol{J}\boldsymbol{\omega}^*_p + 2\dot{\boldsymbol{\omega}}_p^{*\times}\boldsymbol{J}\dot{\boldsymbol{\omega}}^*_p + \boldsymbol{\omega}_p^{*\times}\boldsymbol{J}\ddot{\boldsymbol{\omega}}^*_p\big).
\end{aligned}
\end{equation}
These expressions require up to the third derivative of $\boldsymbol{v}_p^*(t)$ and $\boldsymbol{\omega}_p^*(t)$, and are therefore computed analytically for smooth polynomial or trigonometric trajectory parameterizations.

The tension on each cable is then allocated to achieve the required wrench. Rewriting Eq.~\eqref{eq: wrench_definition}, 
\begin{equation}
\boldsymbol{A}_p \hat{\boldsymbol{f}}_T = \begin{bmatrix}
      c_1\boldsymbol{I}_3 & \dots & c_N\boldsymbol{I}_3\\
        \boldsymbol{Y}_1\boldsymbol{t}_1^{\times}\boldsymbol{R}_{PI}&  \dots & \boldsymbol{Y}_N\boldsymbol{t}_N^{\times}\boldsymbol{R}_{PI}
\end{bmatrix} \begin{bmatrix}
    \boldsymbol{n}_1 \hat{f}_{T,1}\\
    \vdots\\
    \boldsymbol{n}_N  \hat{f}_{T,N}\\
\end{bmatrix} = \bar{\boldsymbol{A}}_p\begin{bmatrix} \hat{\boldsymbol{f}}_{Tvec,1} \\ \vdots \\ \hat{\boldsymbol{f}}_{Tvec,N}\end{bmatrix}
\end{equation}
where $\hat{\boldsymbol{f}}_{Tvec,j}\in\mathbb R^3$ is the mass-normalized tension vector in the direction of the $j^{th}$ cable.

\begin{lemma}\label{lem: config_req}
If all the tether points $\boldsymbol{t}_j$ satisfy the following conditions:
\begin{enumerate}
    \item $\exists\, a_j > 0$, $j=1,\dots,N$, such that $\sum_{i=1}^N a_i c_i m_i = 1$ and $\sum_{i=1}^N a_i c_i m_i \boldsymbol{t}_i = \boldsymbol{0}_{3\times1}$.
    \item $-\sum_{i=1}^N a_i c_i m_i \boldsymbol{t}_i^{\times}\boldsymbol{t}_i^{\times} \succ \boldsymbol{0}$.
\end{enumerate}
Then, the following conclusions are true:
\begin{enumerate}
    \item $\bar{\boldsymbol{A}}_p$ has full row rank;
    \item For a wrench vector $\boldsymbol{w} = [\boldsymbol{w}_1^\top,\, \boldsymbol{w}_2^\top]^\top \in \mathbb{R}^{6}$, the following wrench allocation holds:
\begin{equation}\label{eq: wrench_allocation}
\begin{aligned}
    \hat{\boldsymbol{f}}_{Tvec,j} &= \boldsymbol{n}_j \hat{f}_{T,j} 
    = a_j m_j \big(\boldsymbol{w}_1 + \boldsymbol{R}_{IP}\boldsymbol{t}_j^{\times}\boldsymbol{D}\boldsymbol{w}_2\big),\\
    \boldsymbol{D} 
    &= \Big(\sum_{i=1}^N a_i c_i m_i \boldsymbol{t}_i^{\times}\boldsymbol{t}_i^{\times}\Big)^{-1}\boldsymbol{J}/m_p,
\end{aligned}
\end{equation}
Hence, the total wrench under this allocation equals the required total wrench:
\begin{equation}
\boldsymbol{w}  = \bar{\boldsymbol{A}}_p \begin{bmatrix}
    a_1m_1(\boldsymbol{w}_1 + \boldsymbol{R}_{IP}\boldsymbol{t}_1^{\times}\boldsymbol{D}\boldsymbol{w}_2)\\
    \vdots\\
    a_Nm_N(\boldsymbol{w}_1 + \boldsymbol{R}_{IP}\boldsymbol{t}_N^{\times}\boldsymbol{D}\boldsymbol{w}_2)\\
\end{bmatrix}
\end{equation}
\end{enumerate}

\begin{proof}
    See Appendix~\ref{app: config_req}.
\end{proof}
\end{lemma}

$\boldsymbol{a}_j$ and $\boldsymbol{D}$ can be pre-computed offline for any feasible configurations that satisfy the conditions in Lemma~\ref{lem: config_req}. Hence, the reference tension vector for the $j^{th}$ drone $\hat{\boldsymbol{f}}_{Tvec,j}^*\in\mathbb{R}^3$ is obtained directly from Eq.~\eqref{eq: wrench_allocation} as
\begin{equation}\label{eq: tension_allocation}
    \hat{\boldsymbol{f}}_{Tvec,j}^* = a_j m_j \big(\boldsymbol{w}_1 + \boldsymbol{R}^*_{IP}\boldsymbol{t}_j^{\times}\boldsymbol{D}\boldsymbol{w}_2\big).
\end{equation}
The reference tension magnitude $\hat{f}_{T,j}^*\in\mathbb{R}$ and reference cable direction $\boldsymbol{n}_j^*\in\mathbb{R}^3$ are then recovered as
\begin{equation}
    \hat{f}_{T,j}^* = \|\hat{\boldsymbol{f}}_{Tvec,j}^*\|, \quad 
    \boldsymbol{n}_j^* = \hat{\boldsymbol{f}}_{Tvec,j}^* / \hat{f}_{T,j}^*.
\end{equation}

The reference cable swing velocity $\boldsymbol{v}_j^*$ and swing acceleration $\boldsymbol{z}_j^*$ are obtained from the first and second time derivatives of $\boldsymbol{n}_j^*$. Differentiating Eq.~\eqref{eq: tension_allocation} with respect to time gives the time derivative of the tension vector as
\begin{equation}\label{eq: tension_vec_dot}
    \dot{\hat{\boldsymbol{f}}}_{Tvec,j}^* = a_j m_j \big(\dot{\boldsymbol{w}}_1 + \boldsymbol{R}^*_{IP}{\boldsymbol{\omega}_p^*}^\times\boldsymbol{t}_j^{\times}\boldsymbol{D}\boldsymbol{w}_2 + \boldsymbol{R}^*_{IP}\boldsymbol{t}_j^{\times}\boldsymbol{D}\dot{\boldsymbol{w}}_2\big).
\end{equation}
Decomposing $\dot{\hat{\boldsymbol{f}}}_{Tvec,j}^*$ into components parallel and perpendicular to $\boldsymbol{n}_j^*$ via the projection $\boldsymbol{P}_j = \boldsymbol{I}_3 - \boldsymbol{n}_j^*\boldsymbol{n}_j^{*\top}$, the reference tension rate and direction rate are
\begin{equation}
    \dot{\hat{f}}_{T,j}^* = \boldsymbol{n}_j^{*\top}\dot{\hat{\boldsymbol{f}}}_{Tvec,j}^*, \quad
    \dot{\boldsymbol{n}}_j^* = \frac{\boldsymbol{P}_j\,\dot{\hat{\boldsymbol{f}}}_{Tvec,j}^*}{\hat{f}_{T,j}^*}.
\end{equation}
The second time derivative of $\hat{\boldsymbol{f}}_{Tvec,j}^*$ is similarly obtained from Eq.~\eqref{eq: tension_allocation} as
\begin{equation}\label{eq: tension_vec_ddot}
    \ddot{\hat{\boldsymbol{f}}}_{Tvec,j}^* = a_j m_j \bigg(\ddot{\boldsymbol{w}}_1 + \boldsymbol{R}^*_{IP}\big({\boldsymbol{\omega}_p^*}^\times{\boldsymbol{\omega}_p^*}^\times+{(\dot{\boldsymbol{\omega}}_p^*)}^\times\big)\boldsymbol{t}_j^{\times}\boldsymbol{D}\boldsymbol{w}_2 + 2\boldsymbol{R}^*_{IP}{\boldsymbol{\omega}_p^*}^\times\boldsymbol{t}_j^{\times}\boldsymbol{D}\dot{\boldsymbol{w}}_2+ \boldsymbol{R}^*_{IP}\boldsymbol{t}_j^{\times}\boldsymbol{D}\ddot{\boldsymbol{w}}_2\bigg),
\end{equation}
and the second derivative $\ddot{\boldsymbol{n}}_j^*$ is then computed by differentiating $\dot{\boldsymbol{n}}_j^*$, accounting for the time-varying projection $\boldsymbol{P}_j$:
\begin{equation}
    \ddot{\boldsymbol{n}}_j^* = \frac{\boldsymbol{P}_j\,\ddot{\hat{\boldsymbol{f}}}_{Tvec,j}^* 
    - 2\dot{\hat{f}}_{T,j}^*\dot{\boldsymbol{n}}_j^* 
    -  \boldsymbol{n}_j^*\dot{\boldsymbol{n}}_j^{*\top}\dot{\hat{\boldsymbol{f}}}_{Tvec,j}^*}{\hat{f}_{T,j}^*}.
\end{equation}
Finally, the reference cable kinematics are recovered by projecting onto the horizontal plane via $\boldsymbol{P}_{xy} = \begin{bmatrix}\boldsymbol{I}_2 & \boldsymbol{0}_{2\times1}\end{bmatrix} \in \mathbb{R}^{2\times3}$:
\begin{equation}
    \boldsymbol{r}_j^* = \boldsymbol{P}_{xy}\boldsymbol{n}_j^*, \quad
    \boldsymbol{v}_j^* = \boldsymbol{P}_{xy}\dot{\boldsymbol{n}}_j^*, \quad
    \boldsymbol{z}_j^* = \boldsymbol{P}_{xy}\ddot{\boldsymbol{n}}_j^*.
\end{equation}
The complete reference state and control at time $t$ are then
\begin{equation}
    \boldsymbol{x}^*(t) = \begin{bmatrix}
        \boldsymbol{v}_1^{*\top} & \boldsymbol{v}_2^{*\top} & \boldsymbol{v}_3^{*\top} & 
        \boldsymbol{v}_p^{*\top} & \boldsymbol{\omega}_p^{*\top} & 
        \boldsymbol{r}_1^{*\top} & \boldsymbol{r}_2^{*\top} & \boldsymbol{r}_3^{*\top} & 
        \boldsymbol{x}_p^{*\top} & \mathrm{vec}(\boldsymbol{R}_{IP}^*)^\top
    \end{bmatrix}^\top,
\end{equation}
\begin{equation}
    \boldsymbol{u}^*(t) = \begin{bmatrix}
        \boldsymbol{z}_1^{*\top} & \boldsymbol{z}_2^{*\top} & \boldsymbol{z}_3^{*\top} & 
        \hat{f}_{T,1}^* & \hat{f}_{T,2}^* & \hat{f}_{T,3}^*
    \end{bmatrix}^\top.
\end{equation}
This reference generation procedure is fully analytic and computationally inexpensive, requiring only the evaluation of closed-form expressions at each time step. It is applicable to any sufficiently smooth desired payload trajectory, including the circular and gate-passing trajectories considered in Section~\ref{sec: simulation}.

\subsection{Obstacle Avoidance via Cable Length Control}

The decoupled structure of $\Sigma_l$ from $\Sigma_p$ allows the 
cable length to be controlled independently without affecting the 
payload tracking performance. The variable-length capability is exploited to enable obstacle avoidance capabilities
through constrained environments.

Consider a scenario where the multi-drone system must pass through 
a rectangular gate with a minimum clearance height $h_{min}$ and 
maximum clearance height $h_{max}$, defined in the inertial frame. 
The gate imposes the following height constraints on the system:
\begin{equation}\label{eq: gate_constraints}
    h_{min} + \Delta_{p} \leq \boldsymbol{e}_3^\top\boldsymbol{x}_p,
    \quad
    \boldsymbol{e}_3^\top(\boldsymbol{x}_p + \boldsymbol{R}_{IP}\boldsymbol{t}_j 
    + l_j\boldsymbol{n}_j) \leq h_{max} - \Delta_{q},
\end{equation}
where $\Delta_p > 0$ and $\Delta_q > 0$ are safety margins that 
account for the physical dimensions of the payload and the quadrotors 
along the vertical axis, respectively. The first constraint ensures the payload clears the bottom of the gate, and the second ensures each quadrotor clears the top.

Using the motion planning method in Section~\ref{sec: motion_planning}, a reference trajectory is first generated while ensuring the payload reference height $x_{p,z}^*=\boldsymbol{e}_3^\top\boldsymbol{x}_p^*$ satisfies the first constraint in Eq.~\eqref{eq: gate_constraints} at the point of gate traversal:
\begin{equation}
    x_{p,z}^* = h_{min} + \Delta_p.
\end{equation}
Then, the desired cable length and its rate of change $l_{j,d},\dot{l}_{j,d}$ that satisfies the top clearance constraint can be obtained by isolating $l_j$ from the second constraint in Eq.~\eqref{eq: gate_constraints} and subsequently taking the first derivative, where
\begin{equation}\label{eq: l_desired_gate}
    l_{j,d} = \frac{h_{max} - \Delta_q - 
    \boldsymbol{e}_3^\top(\boldsymbol{x}_p^* + 
    \boldsymbol{R}_{IP}^*\boldsymbol{t}_j)}
    {\boldsymbol{e}_3^\top\boldsymbol{n}_j^*}, \quad
    \dot{l}_{j,d} = -\frac{\boldsymbol{e}_3^\top\big(\boldsymbol{v}_p^*+\boldsymbol{R}_{IP}^*(\boldsymbol{\omega}_p^*)^\times\boldsymbol{t}_j + l_{j,d}\dot{\boldsymbol{n}}^*_j  \big)}{\boldsymbol{e}_3^\top\boldsymbol{n}_j^*}.
\end{equation}
Note that the above quantities can always be computed as $\boldsymbol{e}_3^\top\boldsymbol{n}_j^*>0$ from Assumption~\ref{assum: system}. This expression provides the maximum cable length that keeps each quadrotor below the gate ceiling, given the current cable direction $\boldsymbol{n}_j^*$. 

The desired cable length profile $l_{j,d}(t),\dot{l}_{j,d}(t)$ is computed from Eq.~\eqref{eq: l_desired_gate} along the reference trajectory $\boldsymbol{x}^*$ and tracked by the following Proportional-derivative (PD) cable length controller
\begin{equation}\label{eq: cable_length_PD_controller}
    u_{l,j} = -k_p (l_j-l_{j,d}) - k_d (\dot{l}_j-\dot{l}_{j,d})
\end{equation}
where $k_p$ and $k_d$ are positive control gains. The cable length control are stacked together such that $\boldsymbol{u}_l=[u_{l,1},u_{l,2},u_{l,3}]^\top\in\mathbb R^3$. This approach exploits the additional degree of freedom introduced by variable-length cables to pass through height-constrained environments that would be infeasible for a fixed-length system.

\subsection{Inner-loop Lift Vector Tracking for Quadrotor Attitude Subsystem}

As illustrated in Fig.~\ref{fig: control_framework}, the output from the neural feedback controller and the cable length tracker give the commanded inputs
\begin{equation}
    \begin{bmatrix}
        \tilde{\boldsymbol{z}}_{c}^\top & \hat{\boldsymbol{f}}_{T,c}
    \end{bmatrix}^\top=\boldsymbol{k}_{nn}(\boldsymbol{x},\boldsymbol{x}^*;\theta_{k_1},\theta_{k_2})+\boldsymbol{u}^*, \quad \tilde{\boldsymbol{f}}_{||,c}=\boldsymbol{u}_l, 
\end{equation}
where the desired lift force vector for each quadrotor $\boldsymbol{f}_{L,c,j}$ can be rebuilt from $\tilde{\boldsymbol{z}}_{c},\tilde{\boldsymbol{f}}_{||,c}$ using Eq. \eqref{eq: auxiliary_control_definitions} and \eqref{eq: recover_lift}.

Assuming that each quadrotor is controlled by the magnitude of collective lift force $f_j$ and torque inputs $\boldsymbol{\tau}_j$, the control for each quadrotor is obtained as follows. The lift force magnitude is first computed as $f_j=\|\boldsymbol{f}_{L,c,j}\|$. As the lift force of each quadrotor acts on the z-axis of the $\mathcal{F}_j$, the desired z-axis can be computed as $\boldsymbol{n}_{z,j}=\boldsymbol{f}_{L,c,j}/f_j$. A commanded yaw angle $\psi_j$ is then picked for each quadrotor and the heading vector is defined as $\boldsymbol{n}_{h,j}=[\cos(\psi_j),\sin(\psi_j),0]^\top$. The desired x- and y-axis of $\mathcal{F}_j$, and subsequently $\boldsymbol{R}_{Ij,c}$ can be computed as 
\begin{equation}
    \boldsymbol{n}_{y,j} = \frac{\boldsymbol{n}_{z,j}^\times\boldsymbol{n}_{h,j}}{\|\boldsymbol{n}_{z,j}^\times\boldsymbol{n}_{h,j}\|},\quad \boldsymbol{n}_{x,j} = \boldsymbol{n}_{y,j}^\times\boldsymbol{n}_{z,j},\quad \boldsymbol{R}_{Ij,c}=\begin{bmatrix}
        \boldsymbol{n}_{x,j} & \boldsymbol{n}_{y,j} & \boldsymbol{n}_{z,j}
    \end{bmatrix}.
\end{equation}
By numerical differentiation, the commanded angular velocity $\boldsymbol{\omega}_{j,c}$ and angular acceleration $\dot{\boldsymbol{\omega}}_{j,c}$ can also be obtained, where $\boldsymbol{\omega}_{j,c}=(\boldsymbol{R}_{Ij,c}^\top\dot{\boldsymbol{R}}_{Ij,c})^\vee$. The commanded attitude states for each quadrotor are tracked by adopting a classic geometric controller from \cite{Lee2010}, such that 
\begin{equation}\label{eq: attitude_control}
\boldsymbol{\tau}_j = -k_R\boldsymbol{e}_R -k_\omega\boldsymbol{e}_\omega + \boldsymbol{\omega}_j^\times\boldsymbol{J}_j\boldsymbol{\omega}_j - \boldsymbol{J}_j\big( \boldsymbol{\omega}_j^\times\boldsymbol{R}_{Ij}^\top\boldsymbol{R}_{Ij,c}\boldsymbol{\omega}_{j,c} - \boldsymbol{R}_{Ij}^\top\boldsymbol{R}_{Ij,c}\dot{\boldsymbol{\omega}}_{j,c}\big),
\end{equation}
where $k_R,k_\omega>0$ are constant control gains, $\boldsymbol{e}_R = \frac{1}{2}(\boldsymbol{R}_{Ij,c}^{\top}\boldsymbol{R}_{Ij}-\boldsymbol{R}_{Ij}^\top\boldsymbol{R}_{Ij,c})^\vee\in\mathbb{R}^{3}$ and $\boldsymbol{e}_\omega=\boldsymbol{\omega}_j-\boldsymbol{R}_{Ij}^\top\boldsymbol{R}_{Ij,c}\boldsymbol{\omega}_{j,c}$. An attitude error function is defined as $\Psi(\boldsymbol{R}_{Ij},\boldsymbol{R}_{Ij,c})=\frac{1}{2}\mathrm{Tr}(\boldsymbol{I}_3-\boldsymbol{R}_{Ij,c}^\top\boldsymbol{R}_{Ij})$, where $\mathrm{Tr}(\cdot)$ denotes the matrix trace operator. If the initial attitude error $\Psi\big(\boldsymbol{R}_{Ij}(0),\boldsymbol{R}_{Ij,c}(0)\big)<2$, Proposition 1 of \cite{Lee2010} proves that the zero equilibrium of the tracking errors $\boldsymbol{e}_R$ and $\boldsymbol{e}_\omega$ are exponentially stable, which indicates $\boldsymbol{R}_{Ij}\to\boldsymbol{R}_{Ij,c}$ exponentially. 

\subsection{The closed-loop system}

Under this control framework, the control inputs delivered to the $j^{th}$ quadrotor are the collective lift force $f_j$ and torque inputs $\boldsymbol{\tau}_j$. The true lift force for each quadrotor can be computed as $\boldsymbol{f}_{L,j}=\boldsymbol{R}_{Ij}\boldsymbol{e}_3f_j$. The true control input $\boldsymbol{z}$ and $\hat{\boldsymbol{f}}_{||}$ can therefore be recovered using Eq.~\eqref{eq: normalized_control}, \eqref{eq: projection_matrix}, \eqref{eq: normalized_bot_control_property}. Moreover, the actuator of the $j^{th}$ cable pulley system delivers the required cable tension $f_{T,j} = m_j\hat{f}_{T,c,j}$, where $[\hat{f}_{T,c,1},\, \dots,\,\hat{f}_{T,c,3}]^\top = \hat{\boldsymbol{f}}_{T,c}$. The true control input $\hat{\boldsymbol{f}}_T$ is equal to $\boldsymbol{f}_{T,c}$ as the pulley system for each cable is fully-actuated. 

Therefore, these true control inputs will be substituted into $\Sigma_p$ and $\Sigma_l$ for numerical simulations, forming the complete closed-loop system.

\section{Simulation Results}\label{sec: simulation}

In this section, numerical simulations are conducted to validate the proposed control framework for the three-drone variable-length slung payload system. For the inertial properties of the payload, we have $m_p=1.0$ and $\boldsymbol{J}=\mathrm{diag}(0.6, 0.6,0.8)$. For the three quadrotors, we have $m_1=m_2=m_3=1.5$ and $\boldsymbol{J}_1=\boldsymbol{J}_2=\boldsymbol{J}_3=\mathrm{diag}(0.1,0.1,0.3)$. The tether points on the payload are $\boldsymbol{t}_1=[1,0,0]^\top$, $\boldsymbol{t}_2=[-\cos(\pi/3),-\sin(\pi/3),0]^\top$ and $\boldsymbol{t}_3=[-\cos(\pi/3),\sin(\pi/3),0]^\top$. For $N=3$ drones with equal masses and tether points arranged symmetrically on an equilateral triangle, which is the case according to the definitions of $\boldsymbol{t}_1,\boldsymbol{t}_2,\boldsymbol{t}_3$, conditions 1 and 2 of Lemma~\ref{lem: config_req} are satisfied and the allocation weights simplify to $a_1 = a_2 = a_3 = 1/(3c_j m_j)$. 

The implementation details of the training framework in Section ~\ref{sec: training_payload_subsystem} are as follows. The neural networks $\boldsymbol{K}_1$, $\boldsymbol{K}_2$ and $\boldsymbol{\Theta}$ each consists of two spectrally-normalized fully-connected layers with 128 neurons per layer and $\tanh$ activation functions. A total of $K=8192$ training samples are used and training is conducted for 15 epochs using the \textit{Adam} Optimizer. The hyperparameters are selected as $\lambda=0.5$, $\underline{\mathfrak{m}}=0.1$ and $\overline{\mathfrak{m}}=10$.

The rigid-body payload is commanded to follow a circular trajectory of radius $r_c = 3$ m with angular rate $\omega_c = 0.2$ rad/s, while simultaneously rotating in yaw to remain aligned with the heading direction. A rectangular gate is placed at $(-3, 0)$ m in the $xy$-plane with height constraints $h_{min} = 1.25$ m and $h_{max} = 2.5$ m. The safety margins are set to $\Delta_p = \Delta_q = 0.25$ m. To satisfy the constraints in Eq.~\eqref{eq: gate_constraints}, the payload height profile is designed as
\begin{equation}
    x_{p,z}^*(t) = 0.5\cos(\pi + \omega_c t) + 1,
\end{equation}
where $x_{p,z}^*(\pi/\omega_c)=1.5\geq h_{min}+\Delta_p$ at the time of gate traversal $t=\pi/\omega_c$ to satisfy the floor constraint. The reference payload pose is set to
\begin{equation}\label{eq: ref_traj}
    \boldsymbol{x}_p^*(t)=\begin{bmatrix}r_c\cos(\omega_ct) & r_c\sin(\omega_ct) & 0.5\cos(\pi + \omega_c t) + 1\end{bmatrix}^\top \quad \boldsymbol{R}_{IP}^*(t)=\begin{bmatrix} \cos(\omega_ct) & -\sin(\omega_ct) & 0 \\
    \sin(\omega_ct) & \cos(\omega_ct) & 0 \\
    0 & 0 & 1
    \end{bmatrix}, 
\end{equation}
which yields a smooth and differentiable trajectory for the motion planner in Section~\ref{sec: motion_planning} to generate reference motion. The desired cable lengths are then computed from Eq.~\eqref{eq: l_desired_gate} to satisfy the ceiling constraint. 

\begin{figure}[htbp]
    \centering
    \begin{subfigure}[b]{0.63\textwidth}
        \centering
        \includegraphics[width=\textwidth]{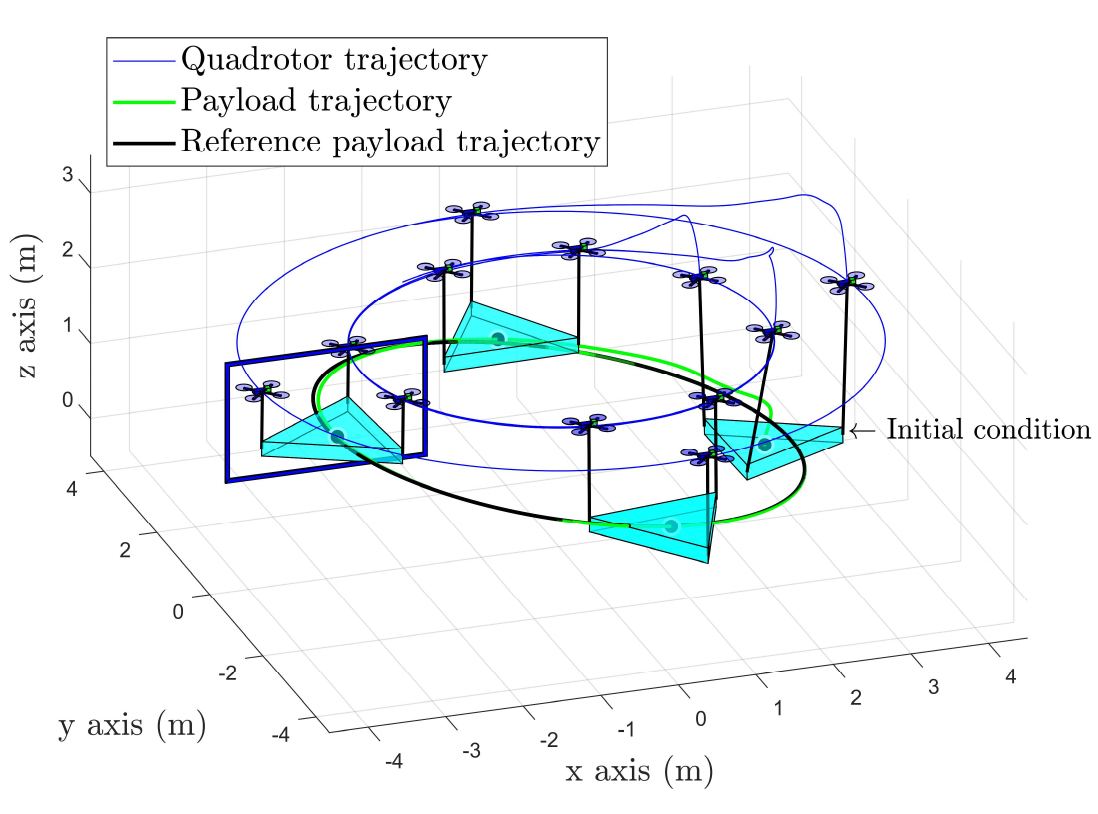}
        \caption{3D trajectories of the payload and the quadrotors during gate traversal.}
        \label{fig: simulation_3d}
    \end{subfigure}
    \hfill
    \begin{subfigure}[b]{0.35\textwidth}
        \centering
        \includegraphics[width=\textwidth]{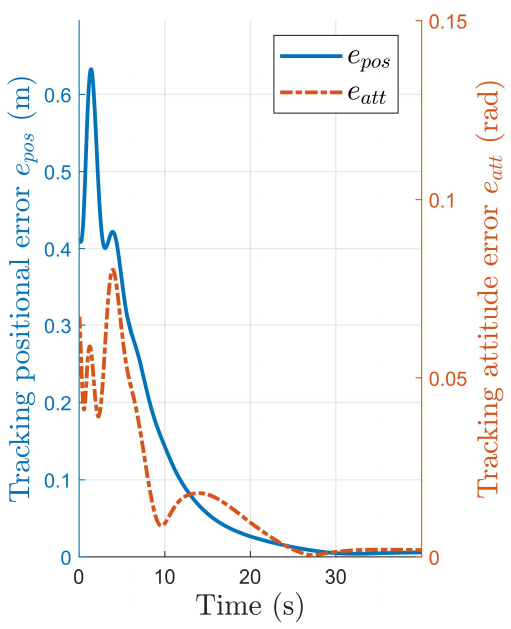}
        \caption{Tracking errors of payload pose.}
        \label{fig: tracking_err}
    \end{subfigure}
    \caption{Numerical simulation results for gate traversal.}
    \label{fig: simulation}
\end{figure}

\begin{figure}[htbp]
    \centering
    \begin{subfigure}[b]{0.54\textwidth}
        \centering
        \includegraphics[width=\textwidth]{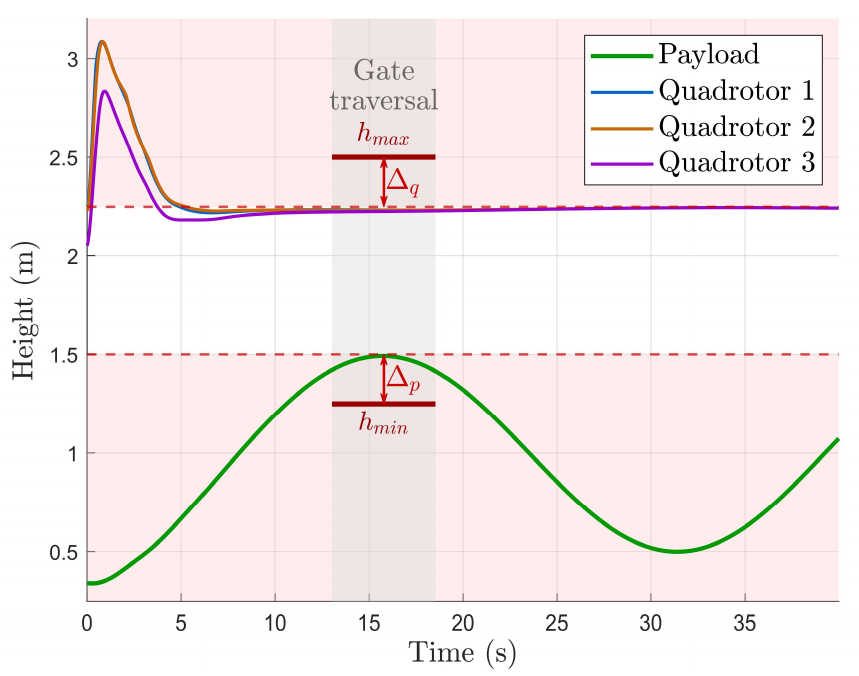}
        \caption{Height of quadrotors and payload during gate traversal.}
        \label{fig: gate_height}
    \end{subfigure}
    \hfill
    \begin{subfigure}[b]{0.44\textwidth}
        \centering
        \includegraphics[width=\textwidth]{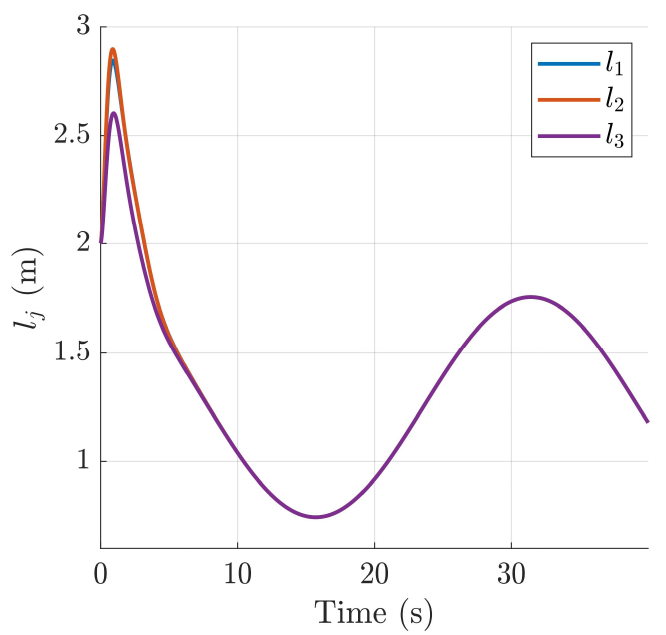}
        \caption{Cable lengths during gate traversal.}
        \label{fig: cable_length}
    \end{subfigure}
    \caption{Demonstration of constraint satisfaction, where cable length shortens during gate traversal, allowing all quadrotors to satisfy the ceiling constraint 
    $h_{max} - \Delta_q$ while the payload satisfies the floor 
    constraint $h_{min} + \Delta_p$.}
    \label{fig: gate}
\end{figure}

\begin{figure}[htbp]
    \centering
    \begin{subfigure}[b]{0.4\textwidth}
        \centering
        \includegraphics[width=\textwidth]{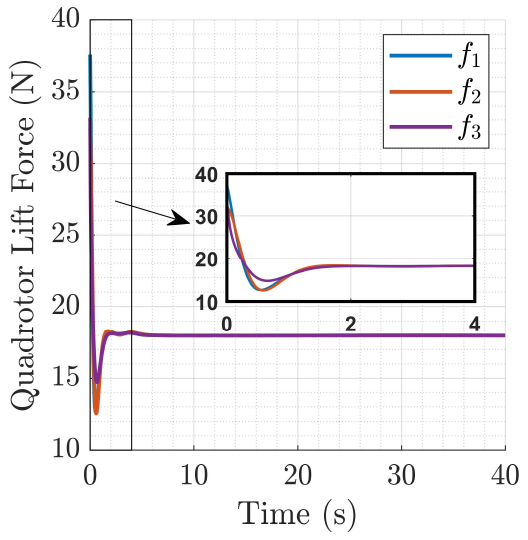}
        \caption{Magnitude of lift from each quadrotor $f_j$.}
        \label{fig: control_input_lift}
    \end{subfigure}
    \hfill
    \begin{subfigure}[b]{0.4\textwidth}
        \centering
        \includegraphics[width=\textwidth]{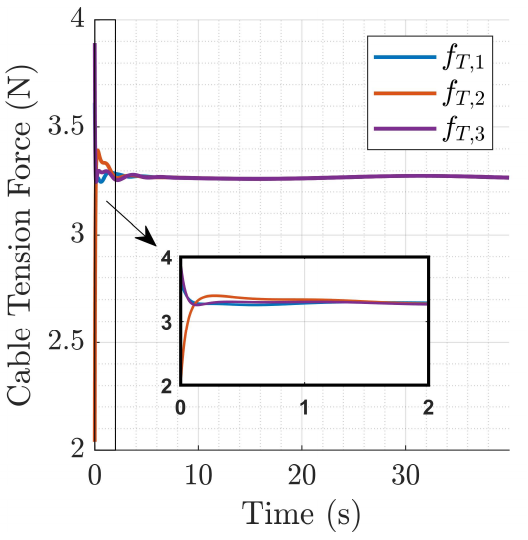}
        \caption{The tension on each cable $f_{T,j}$.}
        \label{fig: control_input_tension}
    \end{subfigure}
    \caption{Control forces in the closed-loop system.}
    \label{fig: control_input}
\end{figure}

The closed-loop system is initialized with an error in the pose and twist of the payload, and the directions and velocities of cables, with respect to $\boldsymbol{x}^*(0)$. The payload trajectory converges to the desired reference trajectory while maintaining stable motion throughout the gate traversal maneuver as illustrated in Fig.~\ref{fig: simulation_3d}. The payload pose tracking errors remain bounded and converge close to zero after the transient response, where the tracking error in payload position and attitude $e_{pos}\in\mathbb{R}$ and $e_{att}\in\mathbb{R}$ are defined as
\begin{equation}
    e_{pos}=\|\boldsymbol{x}_p-\boldsymbol{x}_p^*\|, \quad e_{att} = \cos^{-1}\bigg(\frac{1}{2}\big(\mathrm{Tr}({\boldsymbol{R}_{IP}^*}^{\top}\boldsymbol{R}_{IP})-1\big)\bigg). 
\end{equation}
As shown in Fig.~\ref{fig: tracking_err}, the tracking error converges to near zero within approximately $20$ s, confirming that the neural CCM feedback controller successfully drives $\Sigma_p$ to the reference trajectory in Eq.~\eqref{eq: ref_traj}. 

In Fig.~\ref{fig: gate}, the variable-length cable controller shortens the cables as the system approaches the gate, while the payload reference position is moving upwards. This enables all quadrotors to pass through the gate while remaining within the prescribed safety margin. The ceiling and floor constraints in Eq.~\eqref{eq: gate_constraints} are satisfied during the gate traversal window in the shaded region in Fig.~\ref{fig: gate_height}. The height of each quadrotor stays constant at $h_{max}-\Delta_q$, while the cable lengths are adjusted according to the length profile in Eq.~\eqref{eq: l_desired_gate} as shown in Fig.~\ref{fig: cable_length}. 

Lastly, the control forces applied to the closed-loop system are shown in Fig.~\ref{fig: control_input}, where the magnitudes of lift forces and cable tensions converge to their feed-forward values after transient effects. The implementation code for synthesizing the controller and simulating the closed-loop system is available at our GitHub repository \footnote{\url{https://github.com/loyilok515/SciTech2027}}.

\section{Conclusion}
This paper presented a modular control framework for multi-drone slung payload transportation with variable-length cables and a rigid-body payload. By exploiting the decoupled control-affine structure of the derived equations of motion, a neural CCM controller was trained on the payload subsystem, and an independent cable length controller was designed to exploit the additional degree of freedom for obstacle avoidance.

Future work will focus on experimental validation of the proposed framework. In addition, replacing the PD cable length controller with a learning-based strategy presents a promising direction, enabling the system to autonomously discover length profiles optimized for a broader range of objectives, including active cable swing damping, control effort minimization, and reconfiguration for dynamic obstacle avoidance in unstructured environments.

\section*{Appendix}
\appendix
\section{Proof of Lemma~\ref{lem: B_property}}\label{app: B_property}

From the definitions of $\boldsymbol{n}$ and $\boldsymbol{B}$ in Eq. (\ref{eq: def_cable_vector}) and (\ref{eq: B_property}), the following hold:
\begin{equation}
\boldsymbol{B}^T\boldsymbol{n} = \begin{bmatrix}
\boldsymbol{I}_{2} & -\frac{\boldsymbol{r}}{\sqrt{1-\boldsymbol{r}^T\boldsymbol{r}}}
\end{bmatrix}\begin{bmatrix}
 \boldsymbol{r}\\
 \sqrt{1 - \boldsymbol{r}^T\boldsymbol{r}}
\end{bmatrix} = \boldsymbol{r} - \boldsymbol{r} = \boldsymbol{0}_{2 \times 1}.
\end{equation}
Hence, each row of $\boldsymbol{B}^T$ is orthogonal to $\boldsymbol{n}$. Moreover, since the left $2 \times2$ block of $\boldsymbol{B}^T$ is $\boldsymbol{I}_{2}$, its two rows are linearly independent. Therefore, $\boldsymbol{B}^T$ has full row rank, and its rows form a basis for the 2-dimensional subspace orthogonal to $\boldsymbol{n}$. This proves 1) of Lemma \ref{lem: B_property}. Now let
\begin{equation}
\boldsymbol{P}_B:=   \boldsymbol{B}(\boldsymbol{B}^T\boldsymbol{B})^{-1}\boldsymbol{B}^T.  
\end{equation}
Since the columns of $\boldsymbol{B}$ span the subspace orthogonal to $\boldsymbol{n}$, for any $\boldsymbol{a}_{\bot} \in \mathbb{R}^{3 \times 1}$ satisfying $\boldsymbol{n}^T\boldsymbol{a}_{\bot} = 0$, there exists $\boldsymbol{z}\in \mathbb{R}^{2 \times 1}$ such that
\begin{equation}
  \boldsymbol{a}_{\bot} = \boldsymbol{B}\boldsymbol{z}.  
\end{equation}
Therefore,
\begin{equation}
    \boldsymbol{P}_B \boldsymbol{a}_{\bot} =  \boldsymbol{B}(\boldsymbol{B}^T\boldsymbol{B})^{-1}\boldsymbol{B}^T\boldsymbol{B}\boldsymbol{z} = \boldsymbol{a}_{\bot}.
\end{equation}
On the other hand, if $\boldsymbol{a}_{||}$ is parallel to $\boldsymbol{n}$, then $\boldsymbol{a}_{||} = \alpha \boldsymbol{n}$ for some scalar $\alpha$, and using $\boldsymbol{B}^T\boldsymbol{n}= \boldsymbol{0}_{2 \times 1}$,
\begin{equation}
     \boldsymbol{P}_B \boldsymbol{a}_{||} =  \alpha  \boldsymbol{B}(\boldsymbol{B}^T\boldsymbol{B})^{-1}\boldsymbol{B}^T\boldsymbol{n} = \boldsymbol{0}_{3 \times 1}.
\end{equation}
Thus, for any $\boldsymbol{a}\in \mathbb{R}^{3 \times1}$, decomposed as
\begin{equation}
    \boldsymbol{a} = \boldsymbol{a}_{\bot} + \boldsymbol{a}_{||},
\end{equation}
we obtain
\begin{equation}
    \boldsymbol{P}_B  \boldsymbol{a} = \boldsymbol{a}_{\bot}.
\end{equation}
But the orthogonal projection onto the subspace perpendicular to $\boldsymbol{n}$ is also given by
\begin{equation}
    (\boldsymbol{I}_3 -  \boldsymbol{n} \boldsymbol{n}^T)\boldsymbol{a} = \boldsymbol{a}_{\bot}.
\end{equation}
Since both operations give the same result for every  $\boldsymbol{a}\in \mathbb{R}^{3 \times1}$, it follows that
\begin{equation}
\boldsymbol{B}(\boldsymbol{B}^T\boldsymbol{B})^{-1}\boldsymbol{B}^T  = \boldsymbol{I}_{3} - \boldsymbol{n}\boldsymbol{n}^T.
\end{equation}
This proves 2) of Lemma \ref{lem: B_property}.

\section{Proof of Lemma~\ref{lem: config_req}}\label{app: config_req}

Let vector $\boldsymbol{\zeta} \in \mathbb{R}^{6 \times 1}$:
\begin{equation}
\boldsymbol{\zeta} = \begin{bmatrix}
    \boldsymbol{\theta}\\
    \boldsymbol{q}
\end{bmatrix},
\end{equation}
where $\boldsymbol{\theta}  \in \mathbb{R}^{3 \times 1}$ and $\boldsymbol{q}  \in \mathbb{R}^{3 \times 1}$ satisfy
\begin{equation}
\boldsymbol{\zeta}^T     \bar{\boldsymbol{A}}_p= \boldsymbol{0}_{1 \times 3N}.
\end{equation}
Equivalently, for each $i=1,\dots, N$,
\begin{equation}\label{eq: lemma_aux_1}
c_i    \boldsymbol{\theta}^T + \boldsymbol{q} ^T \boldsymbol{Y}_i \boldsymbol{t}^{\times}_i \boldsymbol{R}_{PI}= \boldsymbol{0}_{1  \times 3}
\end{equation}
Using condition 1, multiplying each equation in (\ref{eq: lemma_aux_1}) by $a_im_i$ and summing over $i$, we have
\begin{equation}
 \boldsymbol{\theta}^T \Big(\sum_{i=1}^N    a_i c_im_i \Big)\boldsymbol{I}_3 +  \boldsymbol{q} ^T m_p \boldsymbol{J}^{-1} \underbrace{ \Big(\sum_{i=1}^N  a_im_ic_i \boldsymbol{t}^{\times}_i \Big)}_{=\boldsymbol{0}_{3 \times 3}} \boldsymbol{R}_{PI} = \boldsymbol{0}_{1  \times 3}
\end{equation}
Note that $\boldsymbol{Y}_i = m_i \boldsymbol{J}^{-1}$. Hence, $ \boldsymbol{\theta} = \boldsymbol{0}_{3 \times 1}$, and each equation in (\ref{eq: lemma_aux_1}) reduces to
\begin{equation}
\boldsymbol{q} ^T \boldsymbol{Y}_i \boldsymbol{t}^{\times}_i \boldsymbol{R}_{PI} = \boldsymbol{y} ^T m_i\boldsymbol{t}^{\times}_i \boldsymbol{R}_{PI} = \boldsymbol{0}_{1  \times 3} \rightarrow  \boldsymbol{y} ^T m_i\boldsymbol{t}^{\times}_i  = \boldsymbol{0}_{1  \times 3}
\end{equation}
where $\boldsymbol{y} = \boldsymbol{J}^{-1}\boldsymbol{q}$. Using conditions 1 and 2, we multiply each equation on the right by $a_ic_i \boldsymbol{t}_{i}^{\times}$, and sum over $i$. This gives:
\begin{equation}
     \boldsymbol{y} ^T \sum_{i=1}^{N}a_ic_im_i\boldsymbol{t}^{\times}_i\boldsymbol{t}^{\times}_i = \boldsymbol{0}_{1  \times 3}
\end{equation}
Since $-\sum_{i=1}^{N}a_ic_im_i\boldsymbol{t}^{\times}_i\boldsymbol{t}^{\times}_i \succ 0$ is invertible according to condition 2, $\boldsymbol{y}= \boldsymbol{0}_{3  \times 1}$ and $\boldsymbol{q} = \boldsymbol{0}_{3  \times 1}$. Hence,  $\boldsymbol{\zeta}^T     \bar{\boldsymbol{A}}_p= \boldsymbol{0}_{1 \times 3N}$ implies $\boldsymbol{\zeta} = \boldsymbol{0}_{6 \times 1}$. Therefore, $ \bar{\boldsymbol{A}}_p$ has full row rank, proving part 1) of Lemma~\ref{lem: config_req}.

According to condition 1, the force channel of the wrench can be recovered as follows:
\begin{equation}
\begin{aligned}
& \begin{bmatrix}
      c_1\boldsymbol{I}_3 & \dots & c_N\boldsymbol{I}_3\\
\end{bmatrix}    \begin{bmatrix}
    a_1m_1(\boldsymbol{w}_1 + \boldsymbol{R}_{IP}\boldsymbol{t}_1^{\times}\boldsymbol{D}\boldsymbol{w}_2)\\
    \vdots\\
    a_Nm_N(\boldsymbol{w}_1 + \boldsymbol{R}_{IP}\boldsymbol{t}_N^{\times}\boldsymbol{D}\boldsymbol{w}_2)\\
\end{bmatrix}\\
& = \sum_{i=1}^N c_ia_im_i(\boldsymbol{w}_1 + \boldsymbol{R}_{IP}\boldsymbol{t}_i^{\times}\boldsymbol{D}\boldsymbol{w}_2) = (\sum_{i=1}^N c_ia_i m_i \boldsymbol{w}_1 ) +  (\sum_{i=1}^N c_ia_i m_i\boldsymbol{R}_{IP}\boldsymbol{t}_i^{\times}\boldsymbol{D}\boldsymbol{w}_2) \\
& = \boldsymbol{w}_1 + \boldsymbol{R}_{IP}\underbrace{ (\sum_{i=1}^N c_ia_im_i\boldsymbol{t}_i)^{\times}}_{=\boldsymbol{0}}\boldsymbol{D}\boldsymbol{w}_2 = \boldsymbol{w}_1
\end{aligned}
\end{equation}
For the torque channel of the wrench, we have the following:
\begin{equation}
\begin{aligned}
 &  \begin{bmatrix}
        \boldsymbol{Y}_1\boldsymbol{t}_1^{\times}\boldsymbol{R}_{PI}&  \dots & \boldsymbol{Y}_N\boldsymbol{t}_N^{\times}\boldsymbol{R}_{PI}
\end{bmatrix}    \begin{bmatrix}
    a_1m_1(\boldsymbol{w}_1 + \boldsymbol{R}_{IP}\boldsymbol{t}_1^{\times}\boldsymbol{D}\boldsymbol{w}_2)\\
    \vdots\\
    a_Nm_N(\boldsymbol{w}_1 + \boldsymbol{R}_{IP}\boldsymbol{t}_N^{\times}\boldsymbol{D}\boldsymbol{w}_2)\\
\end{bmatrix} \\
& = \sum_{i=1}^N  a_i m_i\boldsymbol{Y}_i\boldsymbol{t}_i^{\times}\boldsymbol{R}_{PI}(\boldsymbol{w}_1 + \boldsymbol{R}_{IP}\boldsymbol{t}_i^{\times}\boldsymbol{D}\boldsymbol{w}_2)\\
& = \boldsymbol{J}^{-1}m_p(\sum_{i=1}^N  a_ic_im_i \boldsymbol{t}_i^{\times}\boldsymbol{R}_{PI}\boldsymbol{w}_1) +  \boldsymbol{J}^{-1}m_p(\sum_{i=1}^N  a_i c_i m_i\boldsymbol{t}_i^{\times}\boldsymbol{t}_i^{\times}\boldsymbol{D}\boldsymbol{w}_2)\\
& =  \boldsymbol{J}^{-1}m_p  \underbrace{ (\sum_{i=1}^Na_ic_i m_i\boldsymbol{t}_i)^{\times}}_{=\boldsymbol{0}}\boldsymbol{R}_{PI}\boldsymbol{w}_1 +\boldsymbol{J}^{-1}(\sum_{i=1}^N  a_i c_i m_i\boldsymbol{t}_i^{\times}\boldsymbol{t}_i^{\times})(\sum_{i=1}^Na_ic_im_i\boldsymbol{t}_i^{\times}\boldsymbol{t}_i^{\times})^{-1}\boldsymbol{J}\boldsymbol{w}_2 = \boldsymbol{w}_2.
\end{aligned}    
\end{equation}
This concludes the proof of part 2) of Lemma~\ref{lem: config_req}.

\bibliography{citations}

\end{document}